\documentclass[%
 aip,
 pop,
 amsmath,amssymb,
 reprint,%
]{revtex4-1}

\usepackage{graphicx}% Include figure files
\usepackage{dcolumn}% Align table columns on decimal point
\usepackage{bm}% bold math
\usepackage[utf8]{inputenc}
\usepackage[T1]{fontenc}
\usepackage{mathptmx}
\usepackage{etoolbox}
\usepackage{amsmath}
\usepackage[usenames,dvipsnames]{color} % Pureley needed for my TODO command
\usepackage[normalem]{ulem}
\usepackage{bm}
\usepackage[bbgreekl]{mathbbol}

\newcommand*{\cB}{\mathcal{B}}

\newcommand*{\cE}{\mathcal{E}}

\newcommand*{\cJ}{\mathcal{J}}

\newcommand*{\cZ}{\mathcal{Z}}

\newcommand*{\ep}{\epsilon}

\newcommand*{\B}{\bm{B}}

\newcommand*{\J}{\bm{J}}

\newcommand*{\BDB}{\B\cdot \dl B}

\newcommand*{\JxB}{\bm{J}\times \bm{B}}

\renewcommand*{\u}{\bm{u}}
\newcommand*{\uB}{\lbr \u\cdot \B\rbr}

\newcommand*{\Jpl}{J_{\|}}

\newcommand*{\dl}{\bm{\nabla}}
\newcommand*{\del}{\partial}
\newcommand*{\BD}{\bm{B}\cdot\bm{\nabla}}
\newcommand*{\uD}{\bm{u}\cdot\bm{\nabla}}

\newcommand*{\lbr}{\left(}
\newcommand*{\rbr}{\right)}

\newcommand*\at[2]{\left.#1\right|_{#2}}

\makeatletter
\newsavebox{\@brx}
\newcommand{\llangle}[1][]{\savebox{\@brx}{\(\m@th{#1\langle}\)}%
  \mathopen{\copy\@brx\mkern2mu\kern-0.9\wd\@brx\usebox{\@brx}}}
\newcommand{\rrangle}[1][]{\savebox{\@brx}{\(\m@th{#1\rangle}\)}%
  \mathclose{\copy\@brx\mkern2mu\kern-0.9\wd\@brx\usebox{\@brx}}}
\makeatother

\renewcommand{\revised}{\bgroup\markoverwith{\textcolor{yellow}{\rule[-.5ex]{.1pt}{2.5ex}}}\ULon}

\makeatletter
\def\@email#1#2{%
 \endgroup
 \patchcmd{\titleblock@produce}
  {\frontmatter@RRAPformat}
  {\frontmatter@RRAPformat{\produce@RRAP{*#1\href{mailto:#2}{#2}}}\frontmatter@RRAPformat}
  {}{}
}%
\makeatother
\begin{document}

\preprint{AIP/123-QED}

% Force line breaks with \\
\title[Sengupta et al.]{Periodic Korteweg-de Vries soliton potentials generate quasisymmetric magnetic field strength in a finite plasma-$\beta$ equilibrium }% Force line breaks with \\

\author{W. Sengupta}
 \altaffiliation[Email: ]{wsengupta@princeton.edu}
 \affiliation{ 
Department of Astrophysical Sciences, Princeton University, Princeton, NJ, 08543%\\This line break forced with \textbackslash\textbackslash
}
\author{R. Madan}
 %\altaffiliation[Email: ]{rmadan@pppl.gov}%Lines break automatically or can be forced with \\
 \affiliation{ 
Department of Astrophysical Sciences, Princeton University, Princeton, NJ, 08543%\\This line break forced with \textbackslash\textbackslash
}
\affiliation{%
Princeton Plasma Physics Laboratory, Princeton, NJ, 08540%\\This line break forced% with \\
}%

\author{S. Buller}
 %\altaffiliation[Email: ]{sb0095@princeton.edu }%Lines break automatically or can be forced with \\
\affiliation{ 
Department of Astrophysical Sciences, Princeton University, Princeton, NJ, 08543%\\This line break forced with \textbackslash\textbackslash
}
\author{N. Nikulsin}
 %\altaffiliation[Email: ]{nnikulsin@princeton.edu}%Lines break automatically or can be forced with \\
 \affiliation{ 
Department of Astrophysical Sciences, Princeton University, Princeton, NJ, 08543%\\This line break forced with \textbackslash\textbackslash
}
\affiliation{
Stellarator Theory Department, Max Plank Institute for Plasma Physics, 17491 Greifswald, Germany
}

\author{E.J. Paul}
 %\altaffiliation[Email: ]{ejp2170@columbia.edu }%Lines break automatically or can be forced with \\
 \affiliation{ 
Columbia University, New York, NY 10027, USA
}

\author{R. Nies}
 %\altaffiliation[Email: ]{rnies@princeton.edu}%Lines break automatically or can be forced with \\
 \affiliation{ 
Department of Astrophysical Sciences, Princeton University, Princeton, NJ, 08543%\\This line break forced with \textbackslash\textbackslash
}
\affiliation{%
Princeton Plasma Physics Laboratory, Princeton, NJ, 08540%\\This line break forced% with \\
}%

\author{A.A. Kaptanoglu}
 %\altaffiliation[Email: ]alankaptanoglu@nyu.edu}
\affiliation{Courant Institute of Mathematical Sciences, New York University, New York, NY 10012, USA}

\author{S.R. Hudson}
 %\altaffiliation[Email: ]{shudson@pppl.gov }%Lines break automatically or can be forced with \\
\affiliation{%
Princeton Plasma Physics Laboratory, Princeton, NJ, 08540%\\This line break forced% with \\
}%

\author{A. Bhattacharjee}
% \altaffiliation[Email: ]{amitava@princeton.edu}
 \affiliation{ 
Department of Astrophysical Sciences, Princeton University, Princeton, NJ, 08543%\\This line break forced with \textbackslash\textbackslash
}

\date{\today}% It is always \today, today,

\begin{abstract}
Quasisymmetry (QS) is a hidden symmetry of the magnetic field strength, $B$, that enables effective confinement of charged particles in a fully three-dimensional (3D) toroidal plasma equilibrium. Such equilibria are typically modeled by the ideal magnetohydrostatic (MHS) equation. The nonlinear, overdetermined nature of the quasisymmetric MHS equations severely complicates our understanding of the interplay between 3D shaping, equilibrium properties such as pressure and rotational transform, and $B$. Progress has been made through expansions near the magnetic axis; however, a more comprehensive theory is desirable. Using a combination of analysis and regression on a large dataset of numerically optimized quasisymmetric stellarators, we demonstrate that there is a hidden lower dimensionality of $B$ on a magnetic flux surface with connections to the theory of periodic solitons. We show that $B$ on a flux surface is determined by three or at most four flux functions, each of which is a critical value of the derivative of $B$ along the field line. While being consistent with the near-axis models, our results are global and hold even on the last closed flux surface.

\end{abstract}

\maketitle

\section{Introduction}
The importance of confining trapped particles in a fully three-dimensional (3D) magnetic field was recognized early in the fusion program. This resulted in the development of the concept of omnigeneity \citep{hall1975, Helander2014}, which requires the second adiabatic invariant, or the bounce invariant, $\Jpl$, to be independent of the field line label, implying a zero net radial drift of particles. Quasisymmetry is an important subclass\citep{boozer1981,boozer1983,landremanCatto2012omnigenity, Helander2014} of omnigeneity, which further requires the magnetic field strength $B$ to have a hidden symmetry. 

The hidden nature of the symmetry arises from the fact that only in special coordinates, the Boozer coordinates, does $B$ become independent of one of the Boozer angles. This also makes the Lagrangian of the guiding-center motion\citep{burby2020, Helander2014,boozer1980guidingcenter} admit symmetries, and Noether's theorem thus directly provides an integral of motion that confines particle orbits. Unlike geometry-based continuous symmetries such as axisymmetry or helical symmetry, the construction of a quasisymmetric field and its associated Boozer coordinates cannot be carried out independently. Requiring a magnetic field to satisfy both the 3D MHS equilibrium condition and be quasisymmetric results in an overdetermined set of equations \citep{garrenboozer1991a,garrenboozer1991b,landreman2019,rodriguez2020i}, which further complicates the situation. 

Despite the challenges, considerable progress has recently been made in constructing vacuum fields or MHS equilibria with excellent quasisymmetry through careful numerical optimization.\citep{landreman_paul_22, landremanBullerDrevlak2022}  In addition to numerical work, analytical progress has been made through various expansion schemes which expands in the distance from the magnetic axis (the near-axis expansions NAE), \citep{garrenboozer1991a,garrenboozer1991b,landreman2018a,landreman2018b,jorgeSenguptaLandreman2020QS_direct,rodriguezSengupta2022weakly,rodriguezSengupta2023constructingQS,rodriguez2023MHDstability,rodriguez2022_thesis} or the distance from a flux surface. \citep{sengupta_Paul2021vacuum} or utilize deviations from axisymmetry. \citep{plunk2018,plunk2020NASE,henneberg2019} or a generalized Grad-Shafranov approach. \citep{sengupta_nikulsin_gaur2023QSHBS,nikulsin_sengupta_Jorge_Bhattacharjee_2024asymptotic_GS,nikulsin_sengupta_buller2025compactGS} Although each one of these expansion schemes has its limitations stemming from the various underlying assumptions needed for analytical tractability, they provide valuable insights into how 3D geometry, MHS equilibrium, and quasisymmetric field strength interact to map out the quasisymmetric configuration space.\citep{landreman2022mapping,rodriguez2022phase} The analytical tools have been found to be valuable in diverse areas such as coil design \citep{giuliani2024direct,giuliani2022single,henneberg2019}, plasma stability \citep{rodriguez2023MHDstability,landremanJorge2020magneticWell}, energetic particles \citep{EJpaul2025shear,figueiredo2024energetic}, designing omnigeneous stellarators \citep{plunk2019,jorge2022singlenfpQI,goodman2023constructing,mata2022direct}  and turbulence \citep{jorge2020useNAE_turbulence,zhu2025collisionlessRHflow}, thereby motivating us to dig deeper into the theory of quasisymmetric MHS equilibrium in search of a comprehensive model. 

To develop a model of quasisymmetric MHS equilibrium that is valid throughout the entire plasma volume, it is necessary to address the overdetermination problem. An overdetermined system should not be taken to mean that solutions to the system do not exist. Indeed, overdetermination focuses our attention on special solutions. To make progress on this vexing problem, we return to the fundamental notion of field-line independence on $\Jpl$. Treating the field-line label $\alpha$ as time, we then leverage a key insight that there are certain time-dependent potentials for which the adiabatic invariants are exact \citep{gjaja_Bhattacharjee_1992asymptotics_reflectionless,berry_Howles_1990fake_Airy_reflectionless}. Typically, adiabatic invariants with analytic potentials are not exact since over a period of time they can change by a factor \citep{landau1960mechanics}, which is $\sim \exp{(-1/\ep)}$, where $\ep\ll 1$ is the slowness or the adiabatic parameter. Interestingly, the coefficient in front of the exponentially small term is proportional to the reflection coefficient\citep{keller1991adiabatic_changes,gjaja_Bhattacharjee_1992asymptotics_reflectionless,berry_Howles_1990fake_Airy_reflectionless} of the potential. For a reflectionless potential, for which the reflection coefficient is identically zero, the change in the adiabatic invariant is exactly zero. Reflectionless potentials are intimately connected with solitons and integrable systems \citep{ablowitz_2011_book_nonlinear_waves,novikov1984theory_solitons} and appear naturally in the large wavelength limit of periodic cnoidal waves. Therefore, an $\alpha$ dependent $B$ can still generate an $\alpha$ independent $\Jpl$ if it belongs to the restricted class of soliton potentials. 

Although it is a priori not obvious that quasisymmetric $B$ belongs to such a restricted class of potentials, this is entirely consistent with the second-order NAE calculations, beyond which NAE is not helpful because of overdetermination. In a previous study \citep{sengupta2024periodickortewegdevriessoliton}, we went beyond the validity of NAE by utilizing several different approaches involving perturbative treatments for simple geometries, ideas from soliton theory such as the Painlev\'e test and regression techniques. We demonstrated that vacuum fields with good quasisymmetry fall within the class of soliton potentials. In particular, we showed that on any given flux surface $\psi$, quasisymmetric magnetic field strengths are periodic traveling wave solutions of well-known soliton equations\citep{ablowitz_2011_book_nonlinear_waves,novikov1974periodic_KdV,kamchatnov2000nonlinear}, the Korteweg de-vries (KdV) for generic rotational transform $\iota $ and the Gardner equation for small $\iota$. For generic $\iota$, we obtained an explicit characterization of $B$ in terms of three flux functions that are related to the maximum and minimum $B$ on the surface and the rotational transform. 

In this work, we extend the soliton approach to finite-beta MHS equilibria with preassigned temperature and density profiles and self-consistent bootstrap current \citep{landremanBullerDrevlak2022}. After establishing the fundamental mathematical formalism for describing quasisymmetric MHS equilibrium in Section \ref{sec:basic_QS}, we describe the nonlinear interaction between global 3D geometry, MHS force balance, and field strength using the Darboux frame in Section \ref{sec:Darboux}. We return to the exploration of the connection between solitons and $B$ in Section \ref{sec:surf_QS_solitons}, where we present the mathematical formalism appropriate to establish the connection. Finally, we discuss our results in Section \ref{sec:results}, where we demonstrate that the connection between QS and soliton potentials persists even in the presence of finite plasma pressure and a self-consistent bootstrap current.

\section{Basic formulation of ideal force-balanced quasisymmetric equilibria} 
\label{sec:basic_QS}
The basic plasma equilibrium model is provided by the ideal magnetohydrostatics (MHS) system, which is governed by the equations \citep{freidberg2014idealMHD}
\begin{align}
    \JxB=\dl p(\psi), \;\; \J =\dl\times \B,\;\; \dl \cdot \B=0.
    \label{eq:MHS_force_balance}
\end{align}
Here $\J$ is current, and $p(\psi)$ is the plasma pressure, which will be assumed to be a smooth function of the magnetic flux function $\psi$. The question of the existence of nested flux surfaces in a toroidal volume without continuous symmetry remains open. However, it has been argued that imposing certain additional conditions, such as mirror symmetry \citep{lortz1970,grad1971plasma} and quasisymmetry \citep{rodriguez2021islands,sengupta_nikulsin_gaur2023QSHBS,nikulsin_sengupta_Jorge_Bhattacharjee_2024asymptotic_GS}, can ensure nestedness while supporting a continuous pressure profile. Below, we briefly review the basic formulation of QS, \citep{Helander2014,rodriguez2022_thesis,freidberg2014idealMHD,burby2020,rodriguez2020a} relevant to ideal MHS equilibrium.

Rodriguez et. al. \cite{rodriguez2020a,rodrigGBC} established the so-called ``weak form" of QS,\citep{constantin2021} which, up to a choice of the flux function $\psi$, implies the existence of a divergence-free vector field $\u$ such that
\begin{subequations}
    \begin{align}
    \dl\cdot \u&=0, \label{eq:divu0}\\
     \u\cdot \dl B&=0, \label{eq:udBis0}\\
     \B\times\u &=\dl \psi. \label{eq:Bxu_gradpsi}
\end{align}
\label{eq:u_relabeling_symm}
\end{subequations}

% From \eqref{eq:Bxu_gradpsi} and the divergence-free nature of $\u,\B$ we can show that $\u\cdot\dl$ and $\B\cdot\dl$ commute, i.e.,
% \begin{align}
%     [\u\cdot\dl,\B\cdot\dl]=0,
%     \label{eq:uB_commutator}
% \end{align}
% where $[A_1, A_2]\equiv A_1 A_2-A_2 A_1$ denotes the usual commutator of two operators $A_1,A_2$.

From \eqref{eq:Bxu_gradpsi} it follows that $\u$ must be of the form
\begin{align}
    \u = (F(\psi)\B-\B \times \dl \psi )/B^2,
    \label{eq:QS_vector_u}
\end{align}
which leads to the two-term form of QS \cite{rodriguez2020a}
\begin{align}
    \B\times \dl \psi \cdot \dl B = F(\psi) \BDB.
    \label{eq:2term_form}
\end{align} 
The quantity $F(\psi)$ is related to the rotational transform $\iota$. In a vacuum field with $\B=G_0 \dl \Phi$ and the choice of toroidal flux as $\psi$, \cite{boozer1980guidingcenter,Helander2014}
\begin{align}
    F(\psi)=G_0 q_N, \quad q_N\equiv (\iota -N)^{-1},
    \label{eq:F_psi_vac}
\end{align}
where  $N$ is the helicity of QS, $2\pi G_0/\mu_0$ is the poloidal current outside the surface and $\Phi$ is a toroidal angle. The two-term form of QS (equivalently $\u\cdot \dl B=0$) leads to a traveling wave (TW) solution for $B$ of the form\citep{sengupta_Paul2021vacuum} 
\begin{align}
    |\B|=B(\Phi + (F(\psi)/G_0)\alpha,\psi), \quad F(\psi)/G_0=q_N.
    \label{eq:TW_gen_B}
\end{align}
Thus, the rotational transform and, hence, the magnetic shear enter the definition of QS \eqref{eq:2term_form} in a fundamental way. Furthermore, the TW form of $B$ indicates that in the Boozer coordinate system $(\psi,\vartheta_B,\phi_B)$,
\begin{align}
    \phi_B=\Phi, \quad \vartheta_B = \alpha + (\iota-N)\Phi, \quad |\B|=B(\vartheta_B,\psi)
    \label{eq:Boozer_coord_def},
\end{align}
which shows that the field strength $B$ depends only on a single angle $\vartheta_B$ and $\psi$. Hence, the hidden symmetry manifests in Boozer coordinates such that the scalar field $B(\psi,\vartheta_B)$ is 2D whereas the vector field $\B(\psi,\vartheta_B,\phi_B)$ is fully 3D.

Instead of (2b), strong QS \citep{burby2020} requires that 
\begin{align}
    \J\times \u + \dl \uB=0.
    \label{eq:strongQS_constraint}
\end{align}
There is no longer any gauge freedom, and $\psi$ must be chosen to be the helical flux
\begin{align}
    \psi_h= \int \frac{d\psi_T}{q_N(\psi_T)},
    \label{eq:psi_h_def}
\end{align}
with $\psi_T$ representing the toroidal flux. Strong QS always implies weak QS, and they are equivalent under the MHS force balance condition. Consistency of the strong form of QS \eqref{eq:strongQS_constraint} and MHS \eqref{eq:MHS_force_balance} forces the current $\J$ to be of the form
\begin{align}
    \J = -p'(\psi)\bm{u}-F'(\psi)\B.
    \label{eq:J_form_QSMHS}
\end{align}
Finally, we have the ``triple-product" form of QS \citep{rodriguez2022_thesis,rodriguez2020a} 
\begin{align}
    \dl \psi \times \dl B \cdot \dl \lbr \BDB \rbr=0.
    \label{eq:triple_product}
\end{align}
The two-term and the triple-product form of QS are equivalent under the assumption of irrational $\iota$ and continuity on rational surfaces \citep{Helander2014}. Alternatively, the form
\begin{align}
    \u = (\dl \psi\times \dl B)/\BDB,
    \label{eq:u_contra_form}
\end{align}
is consistent with \eqref{eq:QS_vector_u}, satisfies \eqref{eq:udBis0}, \eqref{eq:Bxu_gradpsi}, while \eqref{eq:divu0} implies \eqref{eq:triple_product}.

\section{Description of quasisymmetric geometry and field strength in the Darboux frame }
\label{sec:Darboux}
The magnetic field geometry in a toroidal confinement device and the magnetic field strength are intimately connected. For example, it is well-known that in a large-aspect-ratio tokamak, the leading-order magnetic field strength is proportional to $1/R_0(1+x/R_0)$, where $R_0,x$ are the radius of curvature and the distance from the tokamak axis, respectively. Similarly, from the NAE, we find that $B/B_0\approx 1- \kappa x$, where $\kappa, x$ are, respectively, the curvature and distance from the stellarator axis. Thus, when one imposes the QS constraint on the field strength, one is simultaneously imposing a constraint on both the curvature and the geometry. Geometry is further constrained by the MHS force-balance, as is evident, for example, from the Shafranov shift \citep{freidberg2014idealMHD}. As a result, a quasisymmetric MHS equilibrium is overconstrained, as is evident from the NAE formalism. \citep{garrenboozer1991b,landreman2019,rodriguez2022_thesis} In this Section, we introduce a new formalism to describe the overdetermined quasisymmetric MHS equilibrium in toroidal geometry.

Our approach is based on a formalism developed by Schief \citep{schief2003nested} to describe isodynamic \citep{palumbo1968} MHS equilibrium, where the field strength is constant on a flux surface. Isodynamic magnetic fields are thus a special case of QS, which, unfortunately, cannot be realized in a nonsymmetric toroidal geometry relevant to stellarators.\citep{Helander2014,schief2003nested,palumbo1968}  The approach to isodynamic equilibrium, suitably modified for general QS, is nonetheless worthy of further theoretical investigation, as it provides invaluable clues as to how the magnetic field line geometry and the field strength are nonlinearly coupled by QS and force balance. 

Schief \citep{schief2003nested} leverages the geodesic property\citep{palumbo1968,bishopTaylor1986degenerate,schief2003nested} of isodynamic magnetic fields to motivate the use of geodesic coordinates that simplify the analysis. In the general case where field lines are not geodesics, we demonstrate that the Darboux frame\citep{eisenhart2013treatise}, an analog of the Frenet frame used in surface geometry, can be used to provide a systematic approach to encoding all local information about the 3D magnetic field on a flux surface. The nestedness of these flux surfaces then suggests a way to foliate a toroidal volume. 

The Darboux formalism that we develop in this Section is particularly useful for describing quasipoloidal (QP) symmetry where $B=B(\psi,\phi_B)$. Standard NAE arguments show that QP can not be obtained near the axis. However, away from the axis, an approximate QP can be realized utilizing the Darboux formalism. 

\subsection{The Darboux frame and QS}
\label{sec:Darboux_QS}

We will use the Darboux frame (basis vectors) $\mathbb{T}=(\bm{t},\bm{n},\bm{b})$, where $\bm{t}$ is a unit tangent along the magnetic field line, $\bm{n}=\dl\psi/|\dl\psi|$ is the unit normal to a flux surface $\psi$, and $\bm{b}=\bm{t}\times \bm{n}$ is the unit binormal vector. Note that the frame is aligned with the magnetic field, not the magnetic axis, as is usually done in NAE. Next, we use the standard Clebsch variables $(\psi, \alpha, \ell)$ to set up our coordinates. The Darboux frame equations can be written as 
\begin{align}
    \del_\ell \mathbb{T}= \mathbb{\Lambda}_\ell \mathbb{T}, \quad  \mathbb{\Lambda}_\ell \equiv
    \begin{pmatrix}
        \quad 0 \quad +\kappa_n \quad +\kappa_g\\
        -\kappa_n  \qquad 0\; \quad +\tau_g\\
        -\kappa_g \quad -\tau_g \qquad 0
    \end{pmatrix},
    \label{eq:del_ell_T}
\end{align}
where $\kappa_n,\kappa_g, and \tau_g$ are the normal and geodesic curvatures, and the geodesic torsion, respectively.

The derivatives of $\bm{r}$ that generate the covariant \citep{imbert2024introduction} basis are
 \begin{subequations}
     \begin{align}
     \bm{r}_{,\ell}&=\bm{t} \label{eq:rcl}\\
     \bm{r}_{,\alpha}&=W\bm{t} +V \bm{b}\label{eq:rcalpha}\\
     \bm{r}_{,\psi}&=\lambda \bm{t}+\mu \bm{n}+\nu \bm{b}  \label{eq:rcpsi}.
 \end{align}
 \label{eq:rc_system}
 \end{subequations}
 The forms of \eqref{eq:rcl} and \eqref{eq:rcalpha} follow from the Darboux setup, where $\ell$ is the arclength along the magnetic field. Note that $B\bm{r}_{,\alpha}=\dl\ell\times \dl \psi$ is orthogonal to $\bm{n}$. The various quantities $W,V,\lambda,\mu,\nu$ are related to the elements of the metric tensor $g_{ij}$ for the $(\psi,\alpha,\ell)$ coordinate system, for example, $W=g_{\alpha\ell},\lambda=g_{\ell\psi}$. These quantities and their various derivatives will determine the magnetic field, including its magnitude and geometry.

 Since the Jacobian of the Clebsch\cite{Helander2014,haeseleer_flux_coordinates}  coordinates, $\cJ= \bm{r}_{,\psi}\times \bm{r}_{,\alpha}\cdot \bm{r}_\ell$, is related to the field strength through $\cJ^{-1}=B$, we find that
 \begin{align}
     \mu V =\frac{1}{B}.
     \label{eq:Jac_condn}
 \end{align}

The QS vector $\bm{u}$, given by \eqref{eq:QS_vector_u}, now reads
\begin{align}
    \bm{u}= \frac{F(\psi)}{B}\bm{t}-\frac{|\dl\psi|}{B}\bm{b}
    \label{eq:QS_vec_u_Darboux}
\end{align}
In the Clebsch $(\psi,\alpha,\ell)$ coordinates, the condition $\u\cdot\dl B=0$ then takes the form \cite{freidberg2014idealMHD}
\begin{align}
      \del_\alpha B + H(\psi,\alpha)\del_\ell B=0,
    \label{eq:fund_B_alpha_B_ell_reln}
\end{align}
where $H(\psi,\alpha)$ is independent of $\ell$. It follows from \eqref{eq:fund_B_alpha_B_ell_reln} that
\begin{align}
    H(\psi,\alpha)= -\frac{\at{\del_\alpha B}{\ell}}{\at{\del_\ell B}{\alpha}} = \at{\lbr \frac{\del \ell}{\del \alpha}\rbr}{B}
    \label{eq:H_del_ell_del_alpha}
\end{align}
We note that there is a degree of freedom in choosing the $(\ell,\alpha)$ pair. By transforming to the coordinates $(\overline{\ell},\alpha)$, where 
\begin{align}
    \overline{\ell}\equiv \ell -\int H(\psi,\alpha )d\alpha,
    \label{eq:lbar_def}
\end{align}
we find that
\begin{align}
    \u\cdot\dl = -\at{\del_\alpha}{\overline{\ell}}, \;\; \BD=B\del_{\overline{\ell}},\;\;
    \label{eq:uD_BD}
\end{align}
The condition \eqref{eq:fund_B_alpha_B_ell_reln} then reduces to
\begin{align}
    \at{\del_\alpha}{\overline{\ell}}B=0 \quad \text{i.e.}\quad B=B(\overline{\ell},\psi). 
    \label{eq:dalpha_B_TW_B}
\end{align}
Thus, there exists a special TW frame ($\overline{\ell},\alpha$) in QS where the field strength $B$ is independent of $\alpha$ \citep{freidberg2014idealMHD}. As shown elsewhere\citep{sengupta2024periodickortewegdevriessoliton}, the two TW forms of $B$ given by \eqref{eq:Boozer_coord_def} and \eqref{eq:dalpha_B_TW_B} are equivalent . 

We choose to impose QS using the form \eqref{eq:fund_B_alpha_B_ell_reln}, which is equivalent to $\uD B=0$, where $\uD= -\del_\alpha-H\del_\ell$. The expression for $\uD$ suggests the following form of $\u$, 
\begin{align}
    \bm{u}=-\bm{r}_{,\alpha}-H\bm{r}_{,\ell}=-\lbr W +H\rbr\bm{t}-V\bm{b}.
    \label{eq:QS_vec_u_TW_Darboux}
\end{align}
Comparing \eqref{eq:QS_vec_u_Darboux} and \eqref{eq:QS_vec_u_TW_Darboux} and using \eqref{eq:Jac_condn}, we conclude that
\begin{align}
    W=-H -\frac{F(\psi)}{B}, \quad V=\frac{|\dl\psi|}{B}, \quad \mu=\frac{1}{|\dl\psi|}.
    \label{eq:WVmu_eqs}
\end{align}
The geodesic curvature $\kappa_g$, which can be shown\citep{Helander2014} to be 
\begin{align}
\kappa_g = \frac{\B\times \dl \psi\cdot \dl B}{B^2 |\dl\psi|},
    \label{eq:kappa_g_general_form}
\end{align}
is strongly constrained by QS. For the special case of QP with zero toroidal currents, $\kappa_g=0$ identically, which can be easily seen by employing Boozer coordinates. In this case, $\B$ takes the form $\B=G(\psi)\dl \phi_B + K\dl \psi$, such that $\B\times\dl \psi=G(\psi)\dl\phi_B \times \dl \psi$. Since $B=B(\psi,\phi_B)$ in QP, it follows that $\kappa_g \propto\B\times\dl \psi\cdot \dl B=0$.

Using the two-term form of QS \eqref{eq:2term_form} and \eqref{eq:WVmu_eqs} we find that
\begin{align}
    \kappa_g V = F\frac{\del_\ell B}{B^2}=W_{,\ell}.
    \label{eq:kappa_g_QS_form}
\end{align}

The other scalar quantities $(\lambda, \nu,\kappa_n,\tau_g)$ are to be determined from the commutation of the partial derivatives (see below). Finally, we note that the derivatives along the tangent, normal, and binormal are
\begin{align}
    \bm{t}\cdot \dl &= \del_\ell , \quad \bm{b}\cdot \dl = \frac{1}{V}(\del_\alpha-W \del_\ell)\label{eq:del_along_tnb},\\
    \nonumber
    \bm{n}\cdot \dl &= \frac{1}{\mu}\lbr (\del_\psi -\lambda \del_\ell)-\frac{\nu}{V}(\del_\alpha-W \del_\ell)  \rbr.
\end{align}

\subsection{MHS force-balance}
As discussed earlier, the strong and the weak forms of QS are equivalent when the MHS force-balance condition $\JxB=\dl p(\psi)$ is imposed. From now on, we shall assume $\psi$ to be the helical flux $\psi_h$ as defined in \eqref{eq:psi_h_def}. Crossing $\J$ with $\B$ and using the Clebsch form for $\B=\dl\psi\times \dl \alpha$, the MHS force-balance yields
\begin{align}
    \J \cdot \dl \psi=0, \quad  \J\cdot \dl \alpha = p'(\psi)
    \label{eq:MHS_force_bal}.
\end{align}
The first condition implies that the radial current must vanish. The second condition is the pressure balance condition in the presence of plasma current density. To calculate $\J$, it is convenient to represent $\B$ in the following covariant form
\begin{align}
    \B &= B_\psi \dl \psi +B_\alpha \dl \alpha +B_\ell \dl \ell,
    \label{eq:B_cov}
\end{align}
where $B_\chi=\B\cdot \del_\chi \bm{r}$, with $\chi \in \{\psi,\alpha,\ell\}$. From \eqref{eq:rc_system} it follows that
\begin{align}
    B_\psi =\lambda B, \quad B_\alpha = W B, \quad B_\ell = B.
    \label{eq:Bpsi_alpha_ell_cov}
\end{align}
From the curl of the covariant form of $\B$, we obtain
\begin{align}
    \frac{\J}{B}=\bm{r}_{,\psi}\lbr \del_\alpha B-\del_\ell (W B)\rbr+ \bm{r}_{,\alpha}\lbr \del_\ell(\lambda B)-\del_\psi B\rbr\nonumber\\
    +\bm{r}_{,\ell}\lbr \del_\psi (W B)-\del_\alpha(\lambda B)\rbr.
    \label{eq:J_form}
\end{align}

The zero radial current $\J\cdot \dl \psi=0$ using \eqref{eq:J_form} implies that
\begin{align}
   (\del_\alpha + H\del_\ell)B -B W_{,\ell}-(H+W)B_{,\ell}=0
    \label{eq:radial_current}
\end{align}
which is identically satisfied thanks to the QS constraints on $B$ in \eqref{eq:fund_B_alpha_B_ell_reln} and the form of $W$ given in \eqref{eq:WVmu_eqs}. The pressure balance condition $\J\cdot\dl\alpha=p'$ is satisfied if
\begin{align}
   \del_\ell (\lambda B)=\frac{p'(\psi)}{B}+\del_\psi B,
   \label{eq:pressure_bal_lambda}
\end{align}
which can be written as a magnetic differential equation (MDE) for $\lambda B$
\begin{align}
    \BD (\lambda B)=\del_\psi \lbr p(\psi)+ \frac{B^2}{2}\rbr.
    \label{eq:MDE_lambdaB}
\end{align}
We observe that the right side of \eqref{eq:MDE_lambdaB} is the radial derivative of the total pressure. If we average \eqref{eq:MDE_lambdaB} over the flux surface, the right side will not vanish, implying that $\lambda B$ can not be single-valued. Moreover, the right side is also quasisymmetric, i.e., it depends only on $B, \psi$. Since $\uD$ and $\BD$ commute, the MDE \eqref{eq:MDE_lambdaB} shows that $\lambda$ must itself be quasisymmetric, or equivalently,
\begin{align}
    \lbr \del_\alpha + H\del_\ell \rbr \lambda=0.
    \label{eq:lambda_is_QS}
\end{align}
Imposing the force-balance condition \eqref{eq:MHS_force_bal} and simplifying the expression for the current \eqref{eq:J_form} using the form of $\u$ from \eqref{eq:QS_vec_u_TW_Darboux}  we find that
\begin{align}
    \J &= -p'(\psi)\bm{u}-F' \B + J_B \B \label{eq:JB_is_0}\\
    J_B &=\lbr \del_\psi(WB)-\del_\alpha(\lambda B)-\frac{p' H}{B}+F'\rbr. \nonumber
\end{align}
Comparing \eqref{eq:JB_is_0}  and \eqref{eq:J_form_QSMHS} directly gives $J_B=0$. Upon using the expression for $W$ from \eqref{eq:WVmu_eqs} and the quasisymmetric nature of $\lambda$ \eqref{eq:lambda_is_QS} combined with \eqref{eq:pressure_bal_lambda}, we obtain 
\begin{align}
    J_B= B \del_\psi H(\psi,\alpha).
    \label{eq:JB_Hpsi}
\end{align}
Thus, quasisymmetry combined with MHS force-balance implies that $H=H(\alpha)$. In other words, the TW speed must be the same across the entire volume of nested flux surfaces. Since $H$ is arbitrary and can always be transformed away by a proper choice of $\ell,\alpha$, we shall assume $H=0$ in the remainder. Thus, in a quasisymmetric MHS equilibrium described by the Clebsch covariant form \eqref{eq:B_cov}, each component of $\B$ is quasisymmetric under the MHS force-balance condition. 

We have thus far discussed QS and force-balance. In the following Sections, we discuss the on-surface (involving $\ell,\alpha$ derivatives) and the off-surface ($\psi,\ell$ and $\alpha,\psi$ derivatives) consistency conditions.

\subsection{The on-surface consistency conditions}
\label{sec:on_surf_cons}
The $\alpha$ evolution of the Darboux frame is given by
\begin{align}
    \del_\alpha \mathbb{T}= \mathbb{\Lambda}_\alpha \mathbb{T}, \quad  \mathbb{\Lambda}_\alpha\equiv
    \begin{pmatrix}
        \quad 0 \quad +\alpha_1 \quad +\alpha_2\\
        -\alpha_1  \qquad 0\; \quad +\alpha_3\\
        -\alpha_2 \quad -\alpha_3 \qquad 0
    \end{pmatrix}.
    \label{eq:del_alpha_T}
\end{align}
To determine the functions $\alpha_i, i=1,2,3$, we will use the commutation of the partial derivatives with respect to the ``on-surface" variables $(\ell,\alpha)$ at fixed $\psi$ of the quantities $\bm{r},\mathbb{T}$. 

The first set of on-surface consistency equations follow from equating $\del_\alpha \del_\ell \bm{r}=\del_\alpha \bm{t}$ and $\del_\ell \del_\alpha\bm{r}=\del_\ell(W \bm{t}+V \bm{b})$
\begin{align}
    \alpha_1 \bm{n} +\alpha_2 \bm{b}= \del_\ell(W \bm{t}+V \bm{b}).
\end{align}
Using \eqref{eq:del_ell_T} and \eqref{eq:del_alpha_T} we obtain
\begin{subequations}
    \begin{align}
    W_{,\ell}&=\kappa_g V \label{eq:W_ell}\\
    \alpha_1&= \kappa_n W -\tau_g V\label{eq:alpha1_eq}\\
    \alpha_2&= V_{,\ell}+\kappa_g W \label{eq:alpha2_eq}
\end{align}
\label{eq:r_ell_alpha_commutation}
\end{subequations}
Equation \eqref{eq:W_ell} gives the same expression for $\kappa_g$ as \eqref{eq:kappa_g_QS_form}. 

Next, the commutation of the on-surface derivatives of $\mathbb{T}$ can be expressed as a matrix equation $\mathbb{M}_{\ell\alpha}\mathbb{T}=0$, where
\begin{align}
    \mathbb{M}_{\ell\alpha}\equiv \del_\alpha \mathbb{\Lambda}_\ell -\del_\ell \mathbb{\Lambda}_\alpha + [ \mathbb{\Lambda}_\ell, \mathbb{\Lambda}_\alpha].
    \label{eq:M_onsurf_def}
\end{align}
For nontrivial $\mathbb{T}$ we must have $\mathbb{M}_{\ell\alpha}=0$, which yields after appropriate simplifications using \eqref{eq:r_ell_alpha_commutation},
\begin{widetext}
\begin{subequations}
    \begin{align}
    \alpha_3-\tau_g W &=\frac{1}{\kappa_n}(\del_\ell^2+\kappa_g^2-\tau_g^2)V -\frac{1}{\kappa_n}(\del_\alpha -W\del_\ell)\kappa_g \label{eq:alpha3_eq}\\
    \lbr \del_\alpha - W\del_\ell \rbr \kappa_n &=
    -\lbr \del_\ell \tau_g +2\tau_g\del_\ell \rbr V +\frac{\kappa_g}{\kappa_n}\lbr \lbr \del_\ell^2 +\kappa_n^2+\kappa_g^2-\tau_g^2\rbr V -\lbr \del_\alpha -W\del_\ell\rbr \kappa_g \rbr
    \label{eq:del_alpha_kappa_n_eq}\\
    \lbr \del_\alpha - W\del_\ell \rbr \tau_g &=
    \del_\ell\lbr \frac{1}{\kappa_n}\lbr \del_\ell^2+\kappa_n^2+\kappa_g^2-\tau_g^2\rbr V -\frac{1}{\kappa_n}(\del_\alpha-W\del_\ell)\kappa_g\rbr -\lbr \del_\ell \kappa_n -2\kappa_g\tau_g \rbr V.
    \label{eq:del_alpha_tau_g_eq}
\end{align}
\end{subequations}
\end{widetext}
Equations \eqref{eq:del_alpha_kappa_n_eq} and \eqref{eq:del_alpha_tau_g_eq} describe the $\alpha-$ evolution of the normal curvature and the geodesic torsion.

\subsection{Off-surface consistency conditions}
The $\psi$ evolution of the Darboux frame is
\begin{align}
    \del_\psi \mathbb{T}= \mathbb{\Lambda}_\psi \mathbb{T}, \quad  \mathbb{\Lambda}_\psi\equiv
    \begin{pmatrix}
        \quad 0 \quad +\psi_1 \quad +\psi_2\\
        -\psi_1  \qquad 0\; \quad +\psi_3\\
        -\psi_2 \quad -\psi_3 \qquad 0
    \end{pmatrix}.
    \label{eq:del_psi_T}
\end{align}
First we consider the off-surface consistency conditions of $\bm{r}$. Equating $\del_\psi \del_\ell \bm{r}=\del_\psi \bm{t}$ and $\del_\ell(\del_\psi \bm{r})$ we get
\begin{subequations}
    \begin{align}
    \lambda_{,\ell}&=\kappa_n \mu +\kappa_g \nu \label{eq:lambda_ell},\\
    \psi_1&= \mu_{,\ell} +\kappa_n \lambda -\tau_g \nu \label{eq:psi1_eq},\\
    \psi_2&= \nu_{,\ell} +\kappa_g \lambda +\tau_g \mu \label{eq:psi2_eq}.
\end{align}
\end{subequations}
Similarly, equating $\del_\psi \del_\alpha \bm{r}=\del_\alpha(\del_\psi \bm{r})$ and using the previously derived consistency conditions we get
\begin{subequations}
    \begin{align}
    (\del_\alpha-W \del_\ell)\lambda &=(\del_\psi-\lambda \del_\ell)W + (\nu V_{,\ell}-\nu_{,\ell}V)-2\frac{\tau_g}{B}
    \label{eq:del_alpha_lambda},\\
    (\del_\alpha-W \del_\ell)\mu &=(\alpha_3 -\tau_g W)\nu -V(\psi_3-\tau_g \lambda) \label{eq:del_alpha_mu},\\
    (\del_\alpha-W \del_\ell)\nu &= -(\alpha_3 -\tau_g W)\mu +(\del_\psi-\lambda \del_\ell)V.
\end{align}
\end{subequations}

The commutation of the $\ell,\psi$ and $\psi,\alpha$ derivatives of $\mathbb{T}$ leads to the matrix equations $\mathbb{M}_{\ell\psi}=0=\mathbb{M}_{\psi\alpha}$, where
\begin{align}
    \mathbb{M}_{\ell\psi}&= \del_\ell \mathbb{\Lambda}_\psi-\del_\psi \mathbb{\Lambda}_\ell +[\Lambda_\psi,\Lambda_\ell]\nonumber,\\
    \mathbb{M}_{\psi\alpha}&= \del_\psi \mathbb{\Lambda}_\alpha-\del_\alpha \mathbb{\Lambda}_\psi +[\Lambda_\alpha,\Lambda_\psi]
    \label{eq:M_offsurf_def}
\end{align}
Setting $\mathbb{M}_{\psi\alpha},\mathbb{M}_{\ell\psi}$ to zero exhausts all cross-derivative consistency conditions. Note that all of these equations are not independent of each other. The most important of them are the on-surface equations. Provided they can be satisfied, it should be possible to obtain a local equilibrium solution. 

The equations obtained from 
$ \mathbb{M}_{\psi\alpha}=0$ are 
 \begin{subequations}
        \begin{align}
        \del_\psi \alpha_1 -\del_\alpha \psi_1 +\alpha_3 \psi_2 -\alpha_2 \psi_3 &=0
          \label{eq:dpsi_alpha1_eqn}\\
        \del_\psi \alpha_2 -\del_\alpha \psi_2+\alpha_1 \psi_3 -\alpha_3 \psi_1 &= 0
           \label{eq:dpsi_alpha2_eqn}\\
          \del_\psi \alpha_3 -\del_\alpha \psi_3 +\alpha_2 \psi_1 -\alpha_1 \psi_2 &= 0.
           \label{eq:dpsi_alpha3_eqn}
        \end{align}
    \end{subequations}
    while those from $\mathbb{M}_{\ell\psi}=0$ are
\begin{widetext}
    \begin{subequations}
        \begin{align}
           \kappa_n( \psi_3 -\tau_g \lambda)&=\lbr \del_\ell^2 +\kappa_g^2 -\tau_g^2 \rbr \nu - (\del_\psi -\lambda\del_\ell)\kappa_g +\mu (\kappa_g \kappa_n +\del_\ell \tau_g) +2\mu_{,\ell}\tau_g \label{eq:psi3_eqn}\\
           (\del_\psi-\lambda \del_\ell)\kappa_n &= \lbr \del_\ell^2 +\kappa_n^2 -\tau_g^2 \rbr \mu +\nu (\kappa_g \kappa_n -\del_\ell \tau_g)-2\tau_g \nu_{,\ell}+\kappa_g(\psi_3 -\tau_g \lambda)
           \label{eq:del_psi_kappa_n_eqn}\\
           (\del_\psi-\lambda \del_\ell)\tau_g &= 2\tau_g(\mu \kappa_n +\nu \kappa_g)-\kappa_g \mu_{,\ell}+\kappa_n \nu_{,\ell}+\del_\ell(\psi_3 -\tau_g \lambda),
           \label{eq:del_psi_tau_g_eqn}
        \end{align}
    \end{subequations}
\end{widetext}

\subsection{Applications of the Darboux frame approach}
The Darboux frame approach is exact. A significant simplification is achieved when field lines are geodesics such that $\kappa_g=0$. Then the Darboux frame equations decouple significantly. Such is the case for isodynamic\citep{schief2003nested} $B$, where it can be further shown that the Darboux frame equations reduce to the vortex filament equations, which are well-known soliton equations. The overdetermination caused by the isodynamic problem forces a TW condition on the normal curvature  $\kappa_n$, i.e., $\kappa_n$ must depend only on a linear combination of the two angles on a given flux surface. With the TW condition, one gets a cubic equation for $(\del_\ell \kappa_n^2)^2=\text{cubic}(\kappa_n^2)$, which can be solved in terms of elliptic functions if the three roots of the cubic are provided. Unfortunately, the TW condition is too restrictive and implies \citep{schief2003nested} that fields that are exactly isodynamic throughout the entire volume must be either axisymmetric or helically symmetric; i.e., the symmetry vectors in these cases are well-known Killing vectors of Euclidean space \citep{burby2020}.

The geodesic curvature also vanishes for QP with zero toroidal current magnetic fields, for which the Darboux frame approach is once again suitable. Simplification can also be achieved, at least locally, near regions where $\kappa_g$ is small compared to the inverse averaged major radius. These two applications of the Darboux frame approach are currently being pursued and will be the subject of forthcoming publications. 

Clearly, for arbitrary $\kappa_g$, developing a global theory of quasisymmetric MHS is a daunting task. However, as we will demonstrate in the following, progress can be made if we examine one flux surface at a time. 

\section{Description of quasisymmetric geometry and field strength on a single flux surface }
\label{sec:surf_QS_solitons}

To describe fields on a given flux surface, it is convenient to use the standard Clebsch \cite{Helander2014,haeseleer_flux_coordinates} coordinate system $(\psi,\alpha,\ell)$, such that $\B=\dl \psi\times \dl \alpha$ with flux label $\psi$, field-line label $\alpha$, and $\ell$, the arclength along the magnetic field. We impose suitable toroidal cuts since $\alpha$ is a multi-valued function. In these coordinates, the triple-product form \eqref{eq:triple_product} takes a simple form
\citep{freidberg2014idealMHD}
\begin{align}
  \frac{\del B}{\del \ell} = f(B,\psi).
  \label{eq:basic_BDB_relation}
\end{align}
We note that because $\del_\ell B$ can be both positive and negative, we will assume that $f(B,\psi)$ also depends on $\sigma=\text{sign}(\del_\ell B)$.

In toroidal geometry, $B$ is subject to the following periodic boundary condition \citep{Helander2014}:
\begin{align}
    B(\psi,\alpha, \ell)= B(\psi, \alpha, \ell +L(\psi)).
    \label{eq:B_periodicity_condition}
\end{align}
The period of $B$, $L(\psi)$, is called the connection length and is a quantity of great physical interest \cite{barnes2011critically}. 
%Note that $L$ is constant on a flux surface only in perfect QS.\citep{Helander2014} 

There are, in addition, important integral constraints that follow from the requirements of QS. From the $\del_\ell B$ condition \eqref{eq:basic_BDB_relation}, any integral along the magnetic field line between two points of equal values of $B=B_b$ satisfies
\begin{align}
    \del_\alpha \int_{B\leq B_{b}} d\ell\; F(B;\psi) =0.
    \label{eq:integral_inv_QS_omni}
\end{align}
The field line independence of the second adiabatic invariant $\Jpl$ 
\begin{align}
\Jpl = \oint \sqrt{\cE-B}\;d\ell,
\label{eq:Jpl}
\end{align}
where $\cE$ is related to the particle energy, is a special case of \eqref{eq:integral_inv_QS_omni}. It follows from \eqref{eq:integral_inv_QS_omni} that quasisymmetric $B$ can not have local maxima or minima on a flux surface \citep{Helander2014, landremanCatto2012omnigenity}. 

As we saw in Section \ref{sec:Darboux}, the quasisymmetric MHS system of equations is a highly nonlinear coupled set of overdetermined PDEs. We bypass these difficulties by focusing instead on $B$ such that $\del_\alpha \Jpl=0$ holds identically. Details are provided in our previous work \citep {sengupta2024periodickortewegdevriessoliton}, and here we provide a summary.

We assume the following three properties of $B$. First, physical quantities such as $B$ and $\psi$ will be assumed to be analytic and single-valued functions of their arguments. Second, $B$ must be periodic in $\ell$ on a flux surface with a period $L(\psi)$ consistent with \eqref{eq:B_periodicity_condition}. Third, $\u$ must be such that a nontrivial special frame exists where $B$ is $\alpha$ independent.

The first and the second assumptions are essential requirements for a smooth quasisymmetric MHS equilibrium. In contrast, neither of these assumptions applies \citep{caryShasharina1997omnigenity,plunk2019} to omnigeneity. An alternative form of the third requirement is that $B$ must satisfy the fundamental TW condition \eqref{eq:TW_gen_B} everywhere on the flux surface. Although the third condition is trivially satisfied in systems with Killing vectors where $\u$ and $\B$ are completely decoupled, this is not the case for a fully 3D system. 

To proceed further, we shift to the TW frame where $B$ is $\alpha$ independent, and the $\del_\ell B$ equation can be treated as an ordinary differential equation (ODE) in $\ell$ with period $L(\psi)$. We then analytically continue $B,\ell$ to the complex plane and demand analyticity. The assumed single-valuedness property can break down near \textit{critical singularities}, such as branch points of algebraic or logarithmic nature. Moreover, in contrast to solutions of linear ODEs, whose singularities are determined by those of the coefficients, nonlinear ODEs such as the $\del_\ell B$ equation allow \textit{movable} singularities, which depend on the initial conditions. The standard example is $y'+y^{2}=0$, whose general solution, $y=(x-x_0)^{-1}$, depends on the integration constant $x_0$ determined by the initial condition $y(0)=-x_0^{-1}$. When the critical singularities are fixed, the functions can be made single-valued by defining suitable branch cuts or fixing the Riemann surface. However, the presence of singularities, which are both critical and movable, leads to dense multi-valuedness around movable singularities
of solutions and nonintegrability \cite{conte2012painleve_1century,kruskal2007_Painleve}. The absence of movable critical singularities in the general solution of an ODE is called the \textit{Painlev\'e property} of an ODE, which guarantees maintenance of single-valuedness.

Insisting on the Painlev\'e property and periodicity and using classical results from standard Painlev\'e test and extensions of Malmquist's theorem \citep{hille1997_Complex_ODE,eremenko1982meromorphic}, we find that with suitable variable transformations, we can always reduce the $B$ equation in the complex plane to
\begin{align}
    \cB'(\cZ)^2=P_3(\cB(\cZ)),
    \label{eq:complex_cubic_dBdl}
\end{align}
where $P_3$ is a cubic polynomial. The cubic in $B$ is the analog of the cubic in $\kappa_n^2$ for the isodynamic case. The quartic polynomial can also be mapped to the cubic with a change in coordinates. 

Therefore, the equation \eqref{eq:basic_BDB_relation} can now be expressed as an algebraic ODE
\begin{equation}
    \left( \frac{\partial B}{\partial \ell} \right)^2 = \mathcal{D}(\psi) (B_{max}(\psi)-B)(B-B_{min}(\psi))(B-B_X(\psi)).
    \label{eq:cubic_B_eqn}
\end{equation}
This `cubic' equation can also be derived from the Korteweg-de Vries equation,
\begin{align}
      \del_t B+6 B \del_x B+ \del_x^3 B=0,
    \label{eq:modB_KdV}
\end{align}
under a traveling wave ansatz $B=B(\psi,x-c t)$, where $x,t$ are suitably normalized $\ell,\alpha$. Provided three flux functions $B_M,B_m,B_X$ are specified on a given flux surface, $B$ can be explicitly calculated in terms of elliptic functions. This periodic TW solution of \eqref{eq:modB_KdV}, called a cnoidal wave \citep{ablowitz_2011_book_nonlinear_waves,kamchatnov2000nonlinear}, reduces to the reflectionless potential in the limit of the period $L$ becoming infinite. Near the magnetic axis, we can show that we recover the correct near-axis behavior.

Is the Painlev\'e property necessary and sufficient for a quasisymmetric $B$? It is sufficient for QS since both the triple product form \eqref{eq:triple_product} and the periodicity condition can be satisfied. However, the necessity is not apparent, particularly in light of axisymmetry, where no such stringent restriction on $\del_\ell B$ appears. In particular, in axisymmetry, weak singularities such as $R^m\log{R}^n$, with integer $m,n$ often arise in the description\citep{cerfon2010one_size} of $\psi$  and $B$ in cylindrical coordinates. However, these singularities are fixed and, therefore, not of the critical movable type. Further, the geometry and the $B$ decouple. Thus, all field lines are equivalent, and the TW frame, where $H=0$, is the same for all $\alpha$.

There exist axisymmetric configurations of the types discussed by Hernandes and Clemente \citep{hernandesClemente2009extension} and Bishop and Taylor \citep{bishopTaylor1986degenerate}, where $|\dl\psi|^2$ is a polynomial in $R$. It can be shown that  $(\BDB^2)^2$ is a cubic or a quartic in $B^2$ in these cases and thus possesses the Painlev\'e property . The analysis of these equilibria can be conducted on a surface-by-surface basis. The nested surface condition yields coupled nonlinear ordinary differential equations (ODEs) in the flux coordinate for the roots of cubic or quartic polynomials. However, these equilibria are special where the Grad-Shafranov problem can be integrated using algebraic integrals, thanks to the polynomial structure of $|\dl\psi|^2$. In general, the Painlev\'e property does not apply to axisymmetry and helically symmetric configurations. 

QS close to axisymmetry shows a similar degeneracy. In the large-aspect-ratio $\ep_p\ll 1$ limit, with $\ep_p$ being the inverse aspect ratio, we \citep{sengupta_nikulsin_gaur2023QSHBS} have demonstrated that any tokamak can be deformed into an approximate quasi-axisymmetric stellarator where the QS error will scale as $\ep_p^2$. Interestingly, even in the large aspect ratio limit, a generalized Palumbo-like configuration can be constructed \citep{BrownHBS} such that the magnetic field strength has the Painlev\'e property. These configurations retain qualitative structural similarities to QA equilibria at finite aspect ratio, such as small magnetic shear and sharp ridges on the inboard side where $B$ is maximum. We will consider the near-axisymmetry limit in more detail later in Section \ref{sec:results}.

In contrast, in a generic approximate 3D QS, the geometry and $B$ are strongly coupled. Due to the overdetermined nature, the $\u\cdot\dl B=0$ condition can only be approximately satisfied. Consequently, the TW frame $H=0$ is, in general, different for different field lines. Thus, the $\alpha$ dependence of the singularities of $\del_\ell B$ will vary from field line to field line. Since these movable singularities can be critical, achieving or maintaining the single-valuedness of $B$ will be more challenging. However, when we demand the Painlev\'e property, the moving singularities are not critical, which automatically respects the single-valuedness of $B$ irrespective of the 3D nature of the geometry. Therefore, we can argue that the Painlev\'e property is both necessary and sufficient for obtaining robust QS in a 3D stellarator using numerical optimization methods, where single-valuedness and periodicity of $B$ are enforced directly using a Fourier representation.

\section{Results}
\label{sec:results}
\begin{figure}[b]
   \centering
    \includegraphics[width=0.48\textwidth]{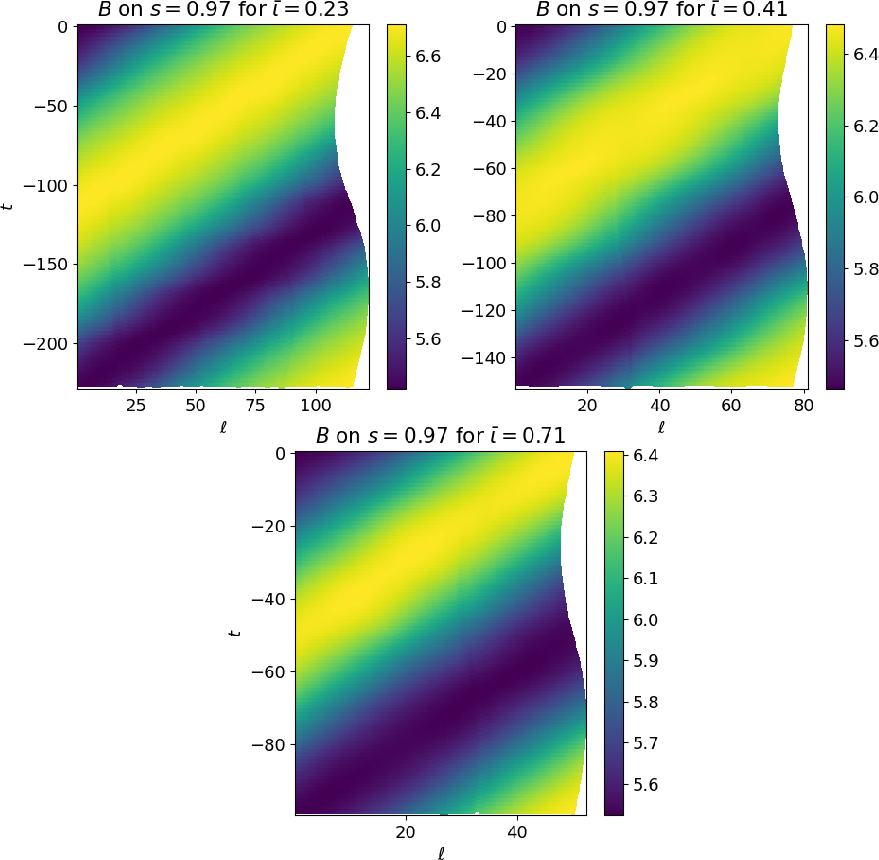}
    \caption{The magnetic field strength plotted on the flux surface parameterized by $(\ell,t)$. As predicted by quasisymmetry, the magnetic field strength displays a traveling waveform.}
    \label{fig:B_wave}
\end{figure}

\begin{figure}
   \centering
    \includegraphics[width=0.48\textwidth]{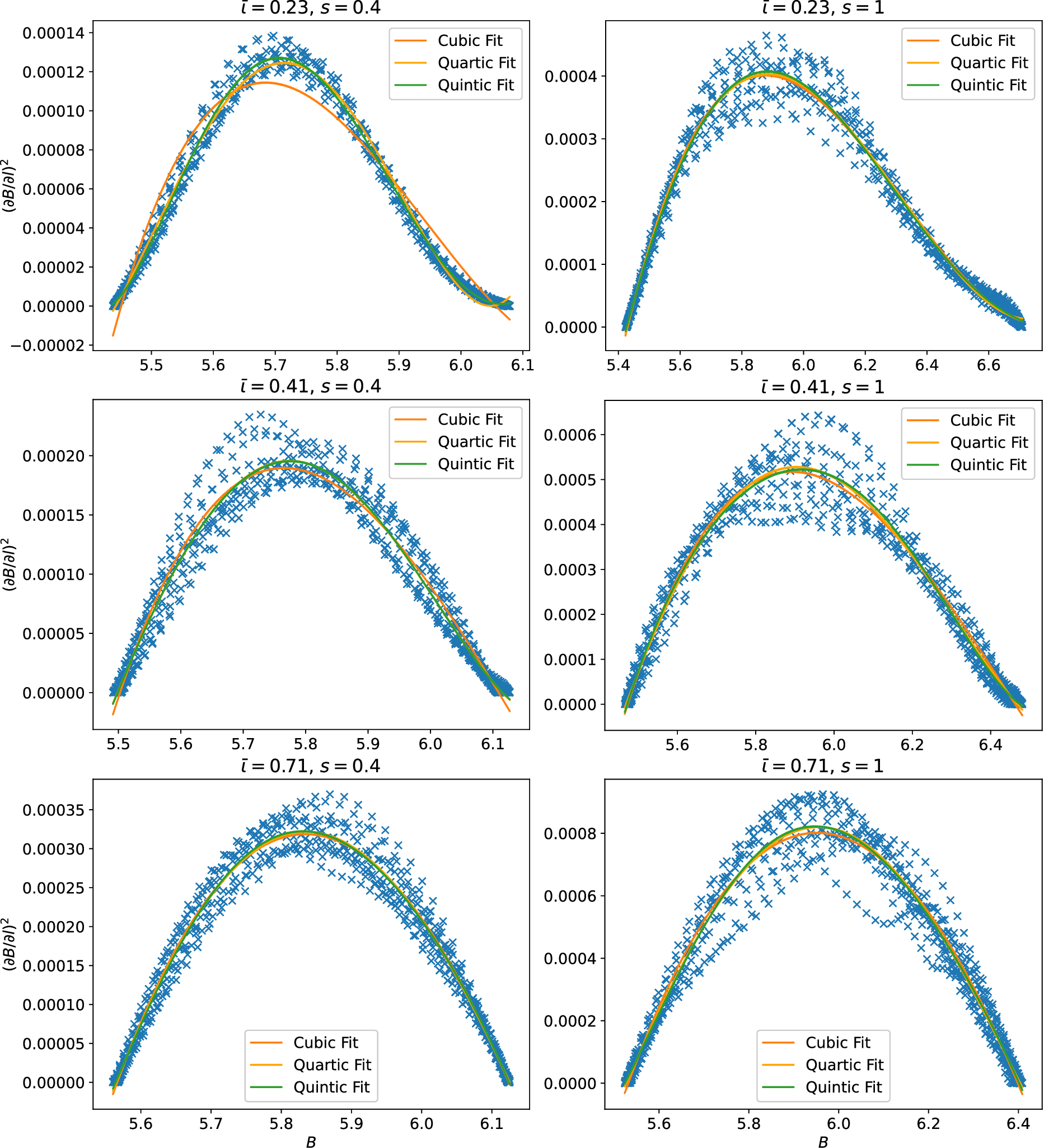}
    \caption{$(\partial_l B)^2$ against $B$ for three different configurations, labeled by their rotational transform, $\iota$, on two different surfaces. For the latter two configurations, a cubic form accurately captures the behavior at both large and small values. At low $\iota$, the cubic is no longer appropriate, and we have to model a quintic to replicate the behavior of the magnetic field strength. As discussed later in Section \ref{sec:near_axisym}, the quintic nature arises from the geometry of the configuration being rather axisymmetric. A more non-axisymmetric geometry at the same value of small $\iota$ requires only a quartic polynomial similar to the vacuum case \citep{sengupta2024periodickortewegdevriessoliton}. }
    \label{fig:dBdl^2vsB}
\end{figure}

% Quasisymmetric field strength satisfies 
% \begin{equation}\label{eq: traveling wave in l alpha}
%     \partial_\ell B + H(\psi,\alpha) \partial_\alpha B=0
% \end{equation}
% where $l$ is the arc length along the field line, $\alpha=\phi - \iota \theta$ with straight field line angles $(\theta,\phi)$, and $H(\psi,\alpha)$ is some function that is independent of $\ell$ but depends on the choice of origin for $\ell$. 

We now present numerical evidence for the main results obtained in the previous section. In the vacuum case, we \citep{sengupta2024periodickortewegdevriessoliton} demonstrated that for general configurations, we obtain the TW form of the KdV equation (cubic). However, for configurations with small rotational transform ($\iota \sim 0.10$), the magnetic field profile is better modeled with a `quartic' equation, which corresponds to the traveling wave ansatz applied to Gardner's equation. Here, we extend these results numerically to configurations with finite beta. In particular, we show that the TW condition \eqref{eq:TW_gen_B}, and the algebraic $(\del_\ell B)^2$ equation (cubic or quartic) are satisfied.

\subsection{Traveling wave}
\label{sec:TW_results}

We define $t=\int^\alpha d\alpha_1 H(\alpha_1)$, such that equation \eqref{eq:fund_B_alpha_B_ell_reln} corresponds to the field strength obeying a traveling waveform, $B=B(\psi,\ell + t)$. %Under such a traveling-wave form, the Korteweg-de Vries equation, which is generically a PDE, reduces to a cubic ODE. 
We verify the traveling wave directly for the finite-$\beta$ configurations we analyze here. We calculate, for a particular flux surface, $H=-\partial_\alpha B/\partial_\ell B$, taking $\ell=0$ at $\phi_B=0$ and considering one field period. Then, with this $H$, we calculate $t$ by  $t=\int^\alpha d\alpha_1 H(\alpha_1)$ and plot $B$ on a flux surface parameterised by $(l,t)$. The results are presented in Figure \ref{fig:B_wave}, where it is indeed verified that even with finite $\beta$, the magnetic field strength appears as a traveling wave with unit speed, as predicted.

\begin{figure}
    \centering
    \includegraphics[width=0.48\textwidth]{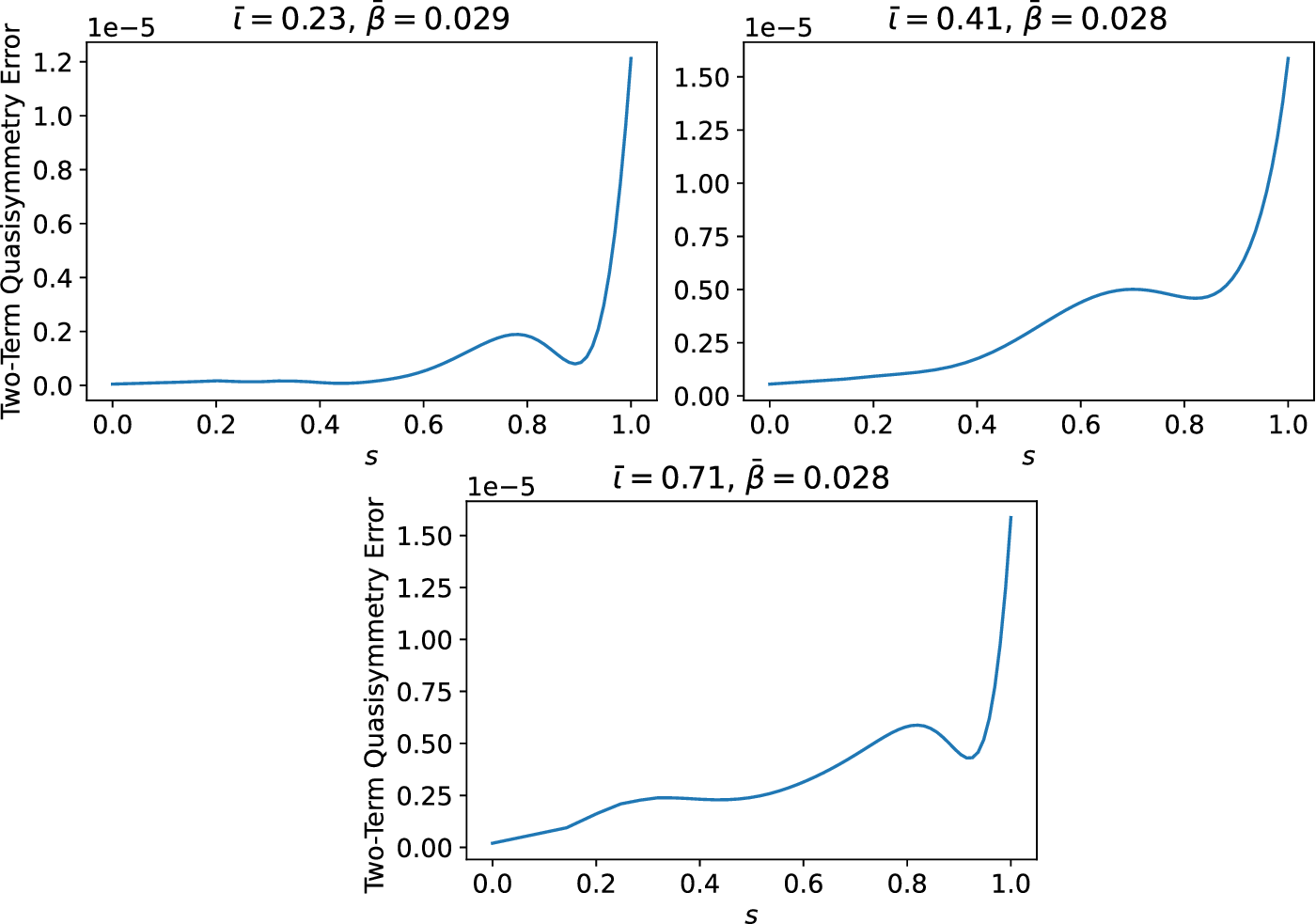}
    \caption{For the three different magnetic field configurations that are considered, we plot the two-term quasisymmetry error.}
    \label{fig:quasisymmetry error profiles}
\end{figure}

\begin{figure}[h]
    \centering
    \includegraphics[width=0.48\textwidth]{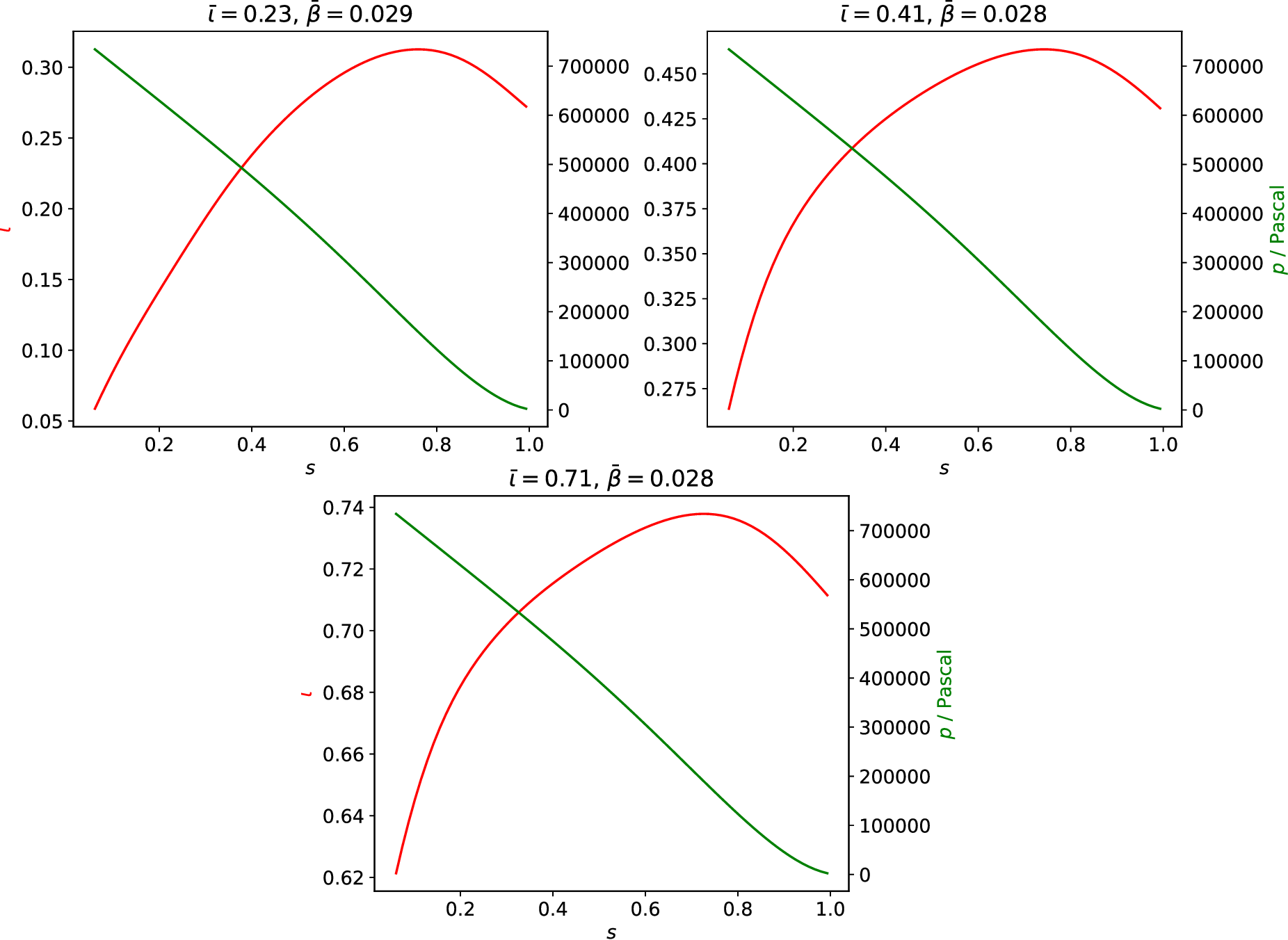}
    \caption{For the three different magnetic field configurations that are considered, we plot the pressure (in green) and iota profiles (in red). $s$ is the normalized toroidal flux.}
    \label{fig:pressure and iota profiles}
\end{figure}

We present three representative configurations from a scan across the mean rotational transform. These configurations were generated by a continuity method\cite{Buller_Landreman_Kappel_Gaur_2025}, by first adding a finite beta to the precise QA magnetic field configuration \cite{landreman_paul_22}, and then varying the target mean rotational transform. As expected, adding a finite beta degraded the quality of the quasisymmetry, which is evident from the spread of points in Figure \ref{fig:dBdl^2vsB}. We also plot the two-term quasisymmetry error, as is usually defined\cite{landreman_paul_22} in {\tt SIMSOPT}, in Figure \ref{fig:quasisymmetry error profiles}. Nonetheless, the configuration is of sufficient quality to reveal distinct trends in the relationship between $B$ and $\partial_\ell B$. The three configurations are labeled by their mean rotational transform ($\bar{\iota}=0.23,0.41,0.71$), which matched their targets. The profiles of rotational transform and pressure are displayed in Figure \ref{fig:pressure and iota profiles}, and their total beta values were $\bar{\beta}=0.029,0.028,0.028$ respectively. 

\subsection{The Effect of Reducing the Axisymmetric Ratio}\label{sec:near_axisym}

As discussed in the previous Section, tokamaks do not need the Painlev\'e property since the symmetry vector is a Killing vector, and geometry completely determines the quasisymmetry vector. Therefore, tokamaks serve as a control in demonstrating that the Painlev\'e property is a truly prominent feature of 3D QS. To this end, we define the axisymmetric fraction ($f_{AS}$) as the summed magnitude of non-axisymmetric modes divided by the magnitude of the axisymmetric mode of the boundary shape: 
\begin{align}
f_{AS}=\frac{%
\sqrt{\sum_{m} \lbr R_{c,m,n=0}^2 + Z_{s,m,n=0}^2 \rbr} %
}{%
\sqrt{\sum_{m,n} \lbr R_{c,m,n}^2 + Z_{s,m,n}^2\rbr} %
}
    \label{eq:AS_fraction}.
\end{align}
By definition, $f_{AS}=1$ for axisymmetric tokamaks. Even the most 3D quasiaxisymmetric configurations tend to have quite high $f_{AS}$ in the range $0.9\leq f_{AS}< 1$. 

We first note that in Figure \ref{fig:dBdl^2vsB}, the $\bar{\iota}=0.23$ case shows that we need a quintic polynomial in $B$ to fit $(\del_\ell B)^2$. However, on closer inspection of the geometry of this configuration, we found that the geometry was essentially axisymmetric $(f_{AS}=0.999)$. Hence, we implemented an additional optimization target that forces the $f_{AS}$ to be below a threshold. By scanning this threshold, we obtain additional $\bar{\iota}=0.23$ configurations with axisymmetric fractions. %in Figure \ref{fig:dBdl^2vsB}, we other configurations with $\bar{\iota}=0.23$ but different axisymmetric fractions. % of $0.995$. %As can be seen from Figure \ref{fig:dBdl^2vsB axisymmetric}    %In Appendix \ref{apdx: reducing axisymmetric ratio}, we illustrate the results of an optimization scan that varies the axisymmetric fraction.

\begin{figure}[b]
   \centering
    \includegraphics[width=0.48\textwidth]{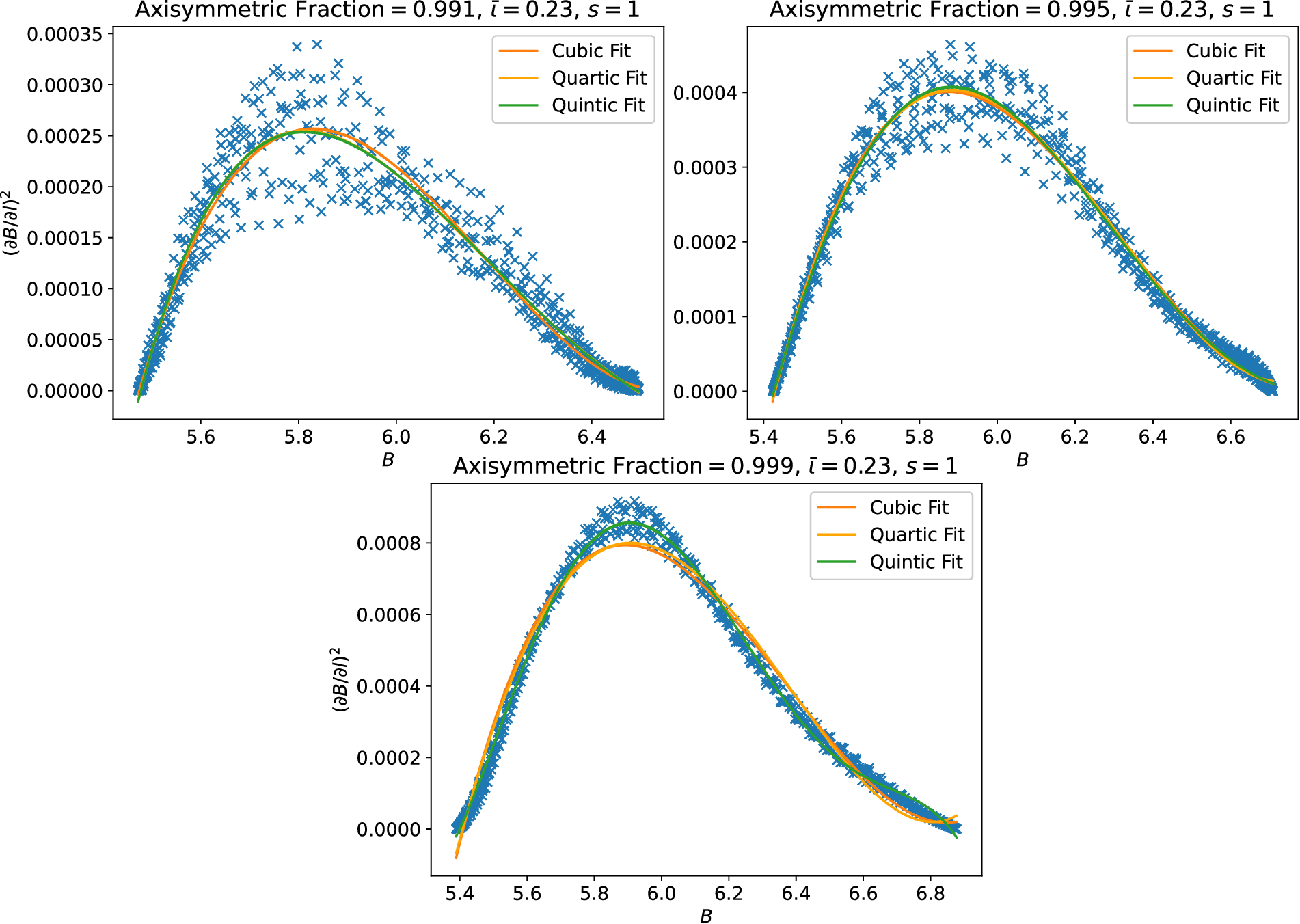}
    \caption{$(\partial_l B)^2$ against $B$ for three representative configurations in the scan of the axisymmetric fraction.}
    \label{fig:dBdl^2vsB axisymmetric}
\end{figure}

\begin{figure}
   \centering
    \includegraphics[width=0.48\textwidth]{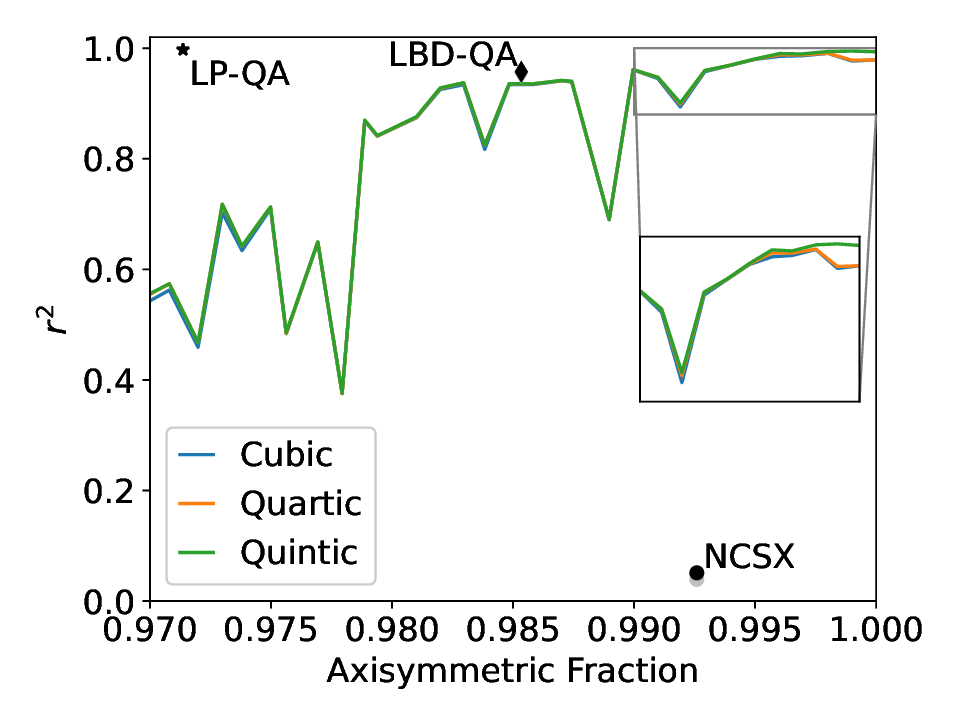}
    \caption{The coefficient of determination on the last closed flux surface for cubic, quartic, and quintic fits across the configurations that span different axisymmetric fractions. Only when we are very close to a tokamak is a cubic no longer a good approximation. Additional QA configurations included as a point of reference: Landreman-Paul QA (LP-QA)\cite{landreman_paul_22}, Landreman-Buller-Drevlak QA (LBD-QA) \cite{landremanBullerDrevlak2022}, and NCSX\cite{zarnstorff2001}.}
    \label{fig:r2 axisymmetric}
\end{figure}

Specifically, we have investigated 31 different $\bar{\iota}=0.23$ configurations that span axisymmetric fractions of $[0.97, 1]$. In Figure \ref{fig:dBdl^2vsB axisymmetric}, we take three of the configurations and plot $(\partial_l B)^2$ against $B$ on the outer surface. We observe that a cubic performs just as well as a quartic or quintic for configurations that are further away from a tokamak; however, for a tokamak, a quintic is required. This trend is supported across all the configurations, as demonstrated in Figure \ref{fig:r2 axisymmetric}, which plots the coefficient of determination of the different polynomial fits on the outer surface across all the configurations. As expected, there is also a trend that, in all polynomial fits, the coefficient of determination is higher for a higher axisymmetric ratio. This is because the quality of quasisymmetry is better for configurations with larger axisymmetric ratios, as can be seen by considering the spread of points in Figure \ref{fig:dBdl^2vsB axisymmetric}. To illustrate the need for quasisymmetry to be well-satisfied, we also show the results of the same fit for NCSX\cite{zarnstorff2001}, the Landreman-Paul QA configuration \cite{landreman_paul_22} and a similar finite-beta QA configuration \cite{landremanBullerDrevlak2022}. As seen in the figure, NCSX has an $r^2$ value close to zero, indicating that it is not well described by a quintic polynomial.

\subsection{The Painlev\'e property shared by finite $\beta$ QS}\label{sec:PP_result}
\begin{figure}[b]
    \centering
    \includegraphics[width=0.48\textwidth]{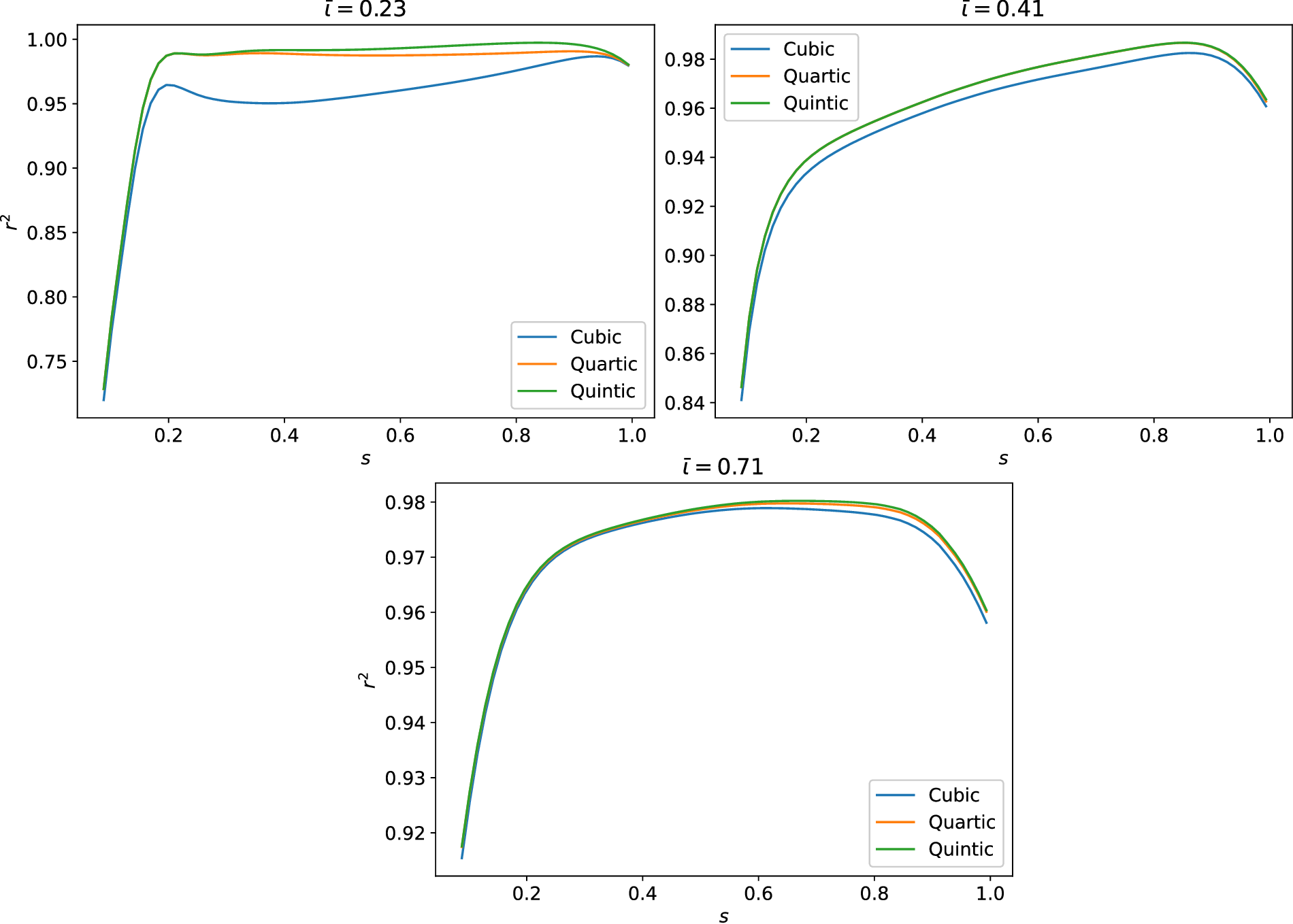}
    \caption{For each surface in each magnetic field configuration, we consider the coefficient of determination for different polynomial fits of $(\partial_l B)^2$ as a function of $B$. $s$ is the normalized toroidal flux.}
    \label{fig:coefficient of determination scan of surf}
\end{figure}

We discussed the near-axisymmetric cases in the previous Section. Now we shall focus on truly 3D configurations. 
The main message is that the cubic generalizes well for the latter two configurations, with $\bar{\iota}=0.41$ and $0.71$ as can be seen from Figure \ref{fig:dBdl^2vsB}, where we show the best fits for a cubic, quartic, and quintic fit plotted against the data for two different surfaces per configuration. For small rotational transform ($\bar{\iota}=0.23$) and sufficiently far from axisymmetry, one needs to only apply a quartic fit to appropriately model the relationship between $(\partial_l B)^2$ and $B$, as can be seen from the first two panels of Figure \ref{fig:dBdl^2vsB axisymmetric}. Also, in Figure \ref{fig:coefficient of determination scan of surf}, we calculate the coefficient of determination, $r^2$, for each polynomial fit. We observe that the cubic model captures the behavior accurately across all surfaces, except for the small rotational transform configuration, where a quartic model outperforms the cubic model significantly. Note that the $r^2$ metric here conflates systematic errors with quasisymmetry errors, which is why the absolute values of $r^2$ are marginally lower for the latter two configurations. Additionally, the drop in performance towards the axis is due to the fact that these configurations were generated by the equilibrium solver {\tt VMEC}, which is known to be inaccurate near the magnetic axis. These results are consistent with previously achieved vacuum results \cite{sengupta2024periodickortewegdevriessoliton}, where a cubic was found, except at low rotational transform, where a quartic was more appropriate.

\begin{figure}
    \centering
    \includegraphics[width=0.48\textwidth]{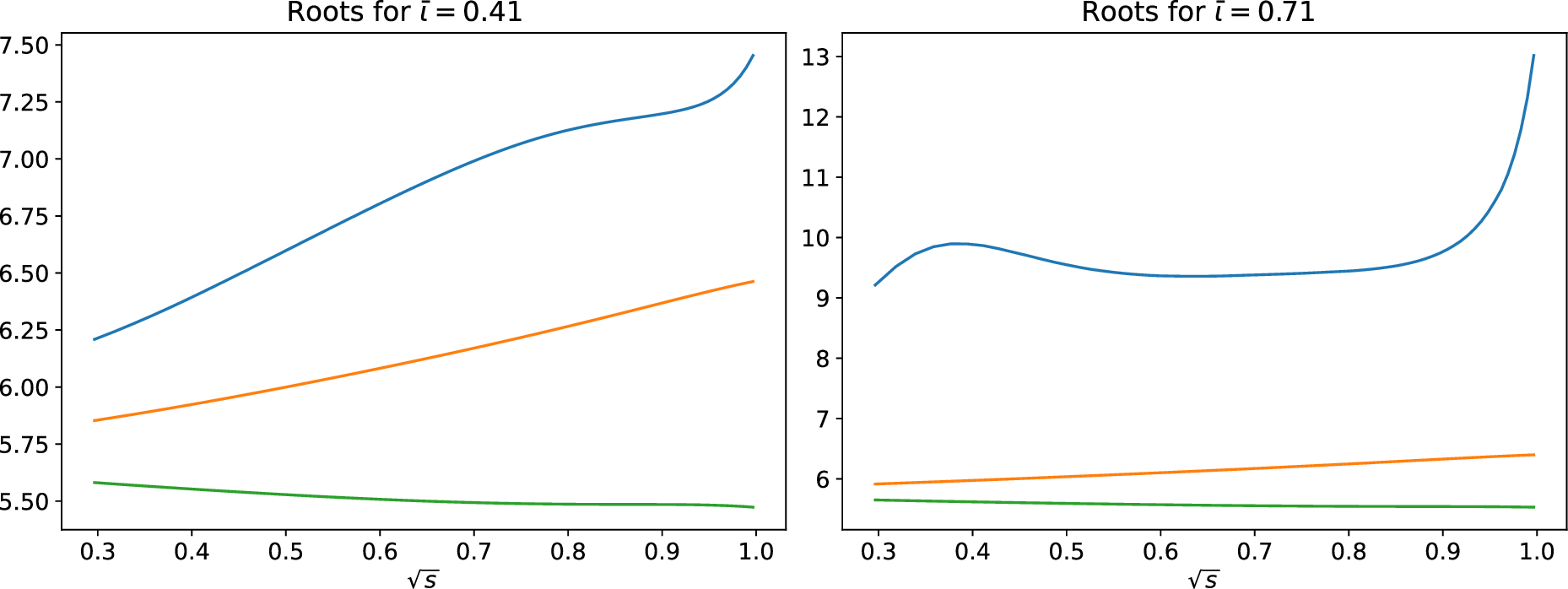}
    \caption{$(\partial_l B)^2$ is fit to a polynomial of $B$ across all flux surfaces -- a cubic in both cases. Shown is the evolution of the roots of the corresponding polynomial across the surfaces; $s$ is the normalized toroidal flux. }
    \label{fig:roots}
\end{figure}

In Figure \ref{fig:roots}, we plot the roots of the best fit polynomial models (quartic for $\bar{\iota}=0.23$, cubic for $\bar{\iota}=0.41,0.71$). In the cubic case for vacuum, it was predicted and verified that $B_{max},B_{min} \propto \sqrt{\psi}$, with $B_{max}$ increasing with $\psi$, $B_{min}$ decreasing with $\psi$, and $B_X$ having a comparatively weak dependence that mirrored the rotational transform profile \citep{sengupta2024periodickortewegdevriessoliton}. For the finite beta $\bar{\iota}=0.41$ configuration, we see that close to the axis, the roots are proportional to $\sqrt{\psi}$, but now $B_{min}$ increases with radius. Towards the last closed flux surface, the relationship between $B_{max}$ and $\sqrt{\psi}$ deviates from linear. For the finite beta $\bar{\iota}=0.71$ configuration, $B_{min}$ and $B_X$ are proportional to $\sqrt{\psi}$, but $B_{max}$ has a more complicated relationship to the radius. %For the $\bar{\iota}=0.23$ case, there is currently no theory for the relationship between roots and flux surfaces; however, we note that there is no sign of overfitting with the quartic, as the roots display robust trends without drastic jumps.

Finally, we note that these configurations were produced with 12 toroidal modes and 16 poloidal modes. We found it necessary to increase the resolution to this higher level, as we encountered unphysical artifacts in the profiles of $(\partial_l B)$ against $B$ at a lower resolution of six toroidal modes and six poloidal modes. Increasing the resolution helped identify these artifacts as unphysical and recovered the low-degree polynomial structure, as is shown in Appendix \ref{apdx: resolution}.

\section{Conclusion}
In conclusion, we have demonstrated that a broad class of excellent quasisymmetric $B$ obtained using standard numerical optimization is generated using periodic soliton potentials. Our previous work \citep{sengupta2024periodickortewegdevriessoliton} treated vacuum fields, and in the present work, we have treated MHS equilibrium with a self-consistent bootstrap current \citep{Buller_Landreman_Kappel_Gaur_2025}.

Employing the Darboux frame, we present the mathematical formalism necessary to describe a generic quasisymmetric MHS equilibrium, which yields a system of nonlinearly coupled overdetermined partial differential equations (PDEs). The Darboux frame formalism is very general and enables analytical progress in cases of quasipoloidal symmetry and those where the geodesic curvature can be neglected, such as the isodynamic case. However, the general QS-MHS system is too complicated to be solved exactly. Instead, we have made progress with a surface-by-surface approach. 

Motivated by the observation that certain special time-dependent potentials connected with reflectionless potentials and their periodic generalization can lead to exact time-invariant integrals, we take a different approach to obtaining constraints on the QS-MHS system. We show that under three very general assumptions, namely, periodicity in $\ell$, analyticity of $B$, and the existence of a nontrivial TW frame, we can motivate the Painlev\'e property for a quasisymmetric $B$, which ensures $B$ stays single-valued. In particular, we show that QS naturally leads to algebraic ODEs for $(\del_\ell B)^2$, which are cubic or quartic in $B$.

\begin{figure}[b]
   \centering
    \includegraphics[width=0.48\textwidth]{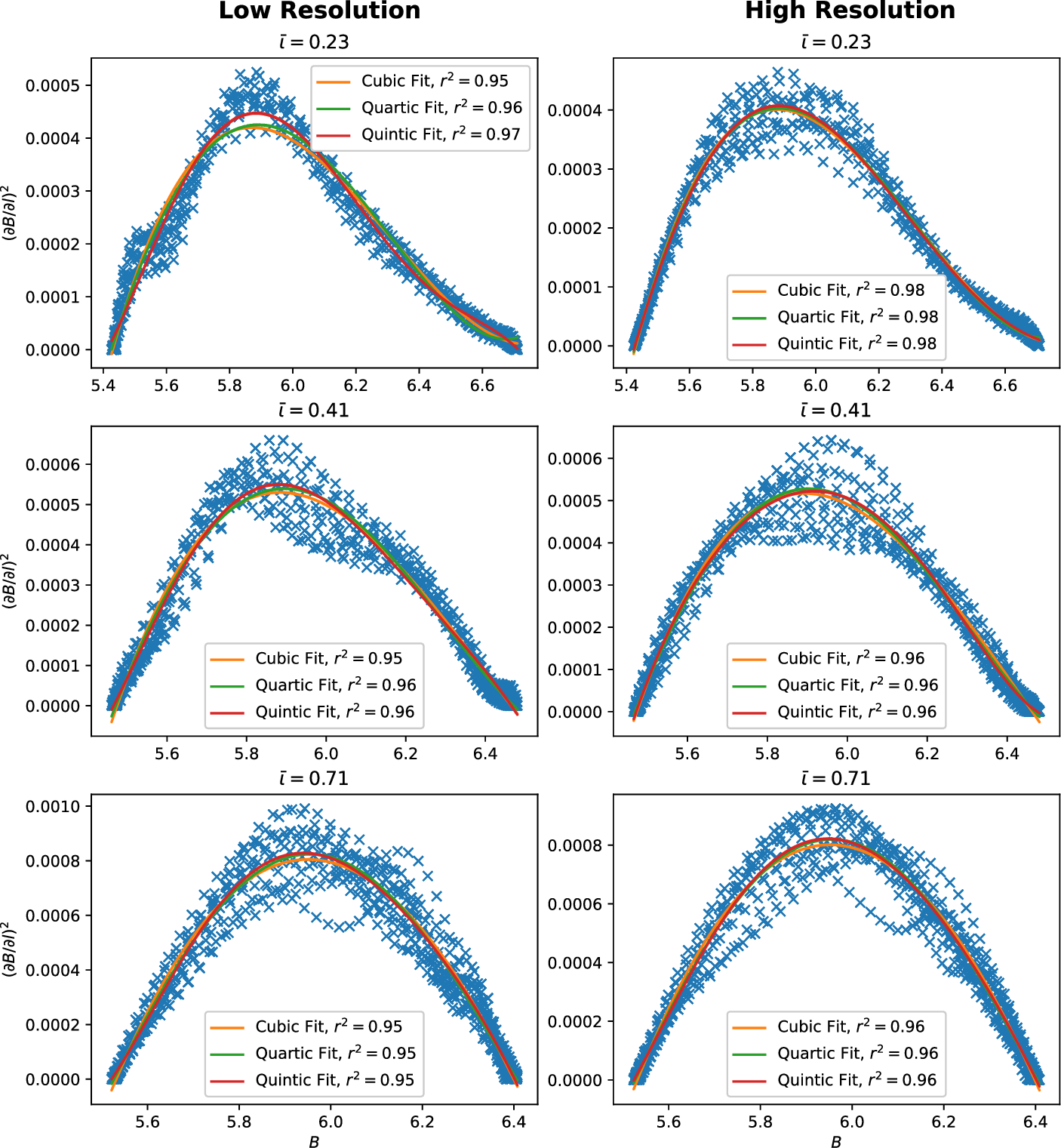}
    \caption{$(\partial_l B)^2$ against $B$ for configurations at three different mean rotational transforms, with their low and high-resolution versions. The low-degree polynomial structure becomes more precise at high resolution.}
    \label{fig:lowres to hires iotascan}
\end{figure}

Using a numerical regression approach, we have demonstrated that the Painlev\'e property holds for finite pressure and bootstrap currents, just as it does in a vacuum. As a control factor, we have looked at the near-axisymmetric limit. We showed \citep{sengupta_nikulsin_gaur2023QSHBS} that, in the large aspect ratio limit, any tokamak can be deformed into an approximately quasi-axisymmetric device, irrespective of the shape of its cross-section. Thus, as we approach the axisymmetric limit, which we formally defined in terms of an axisymmetric fraction going towards unity, we expect deviations from the cubic and quartic nature of $(\del_\ell B)^2$. Restoring sufficient 3D shaping restores the cubic/quartic or the Painlev\'e property. Lastly, the importance of having a good resolution is demonstrated.

\appendix

\section{The Effect of Increasing Resolution}\label{apdx: resolution}
In this Appendix, we demonstrate that a sufficiently high resolution is required to precisely reveal the polynomial structure of the magnetic field strength profile. We compare `low-resolution' configurations, with six toroidal and six poloidal modes, with `high-resolution' configurations, which have 12 toroidal modes and 16 poloidal modes. Both sets of configurations have the same radial resolution, with 75 surfaces.

\begin{figure}[b]
   \centering
    \includegraphics[width=0.48\textwidth]{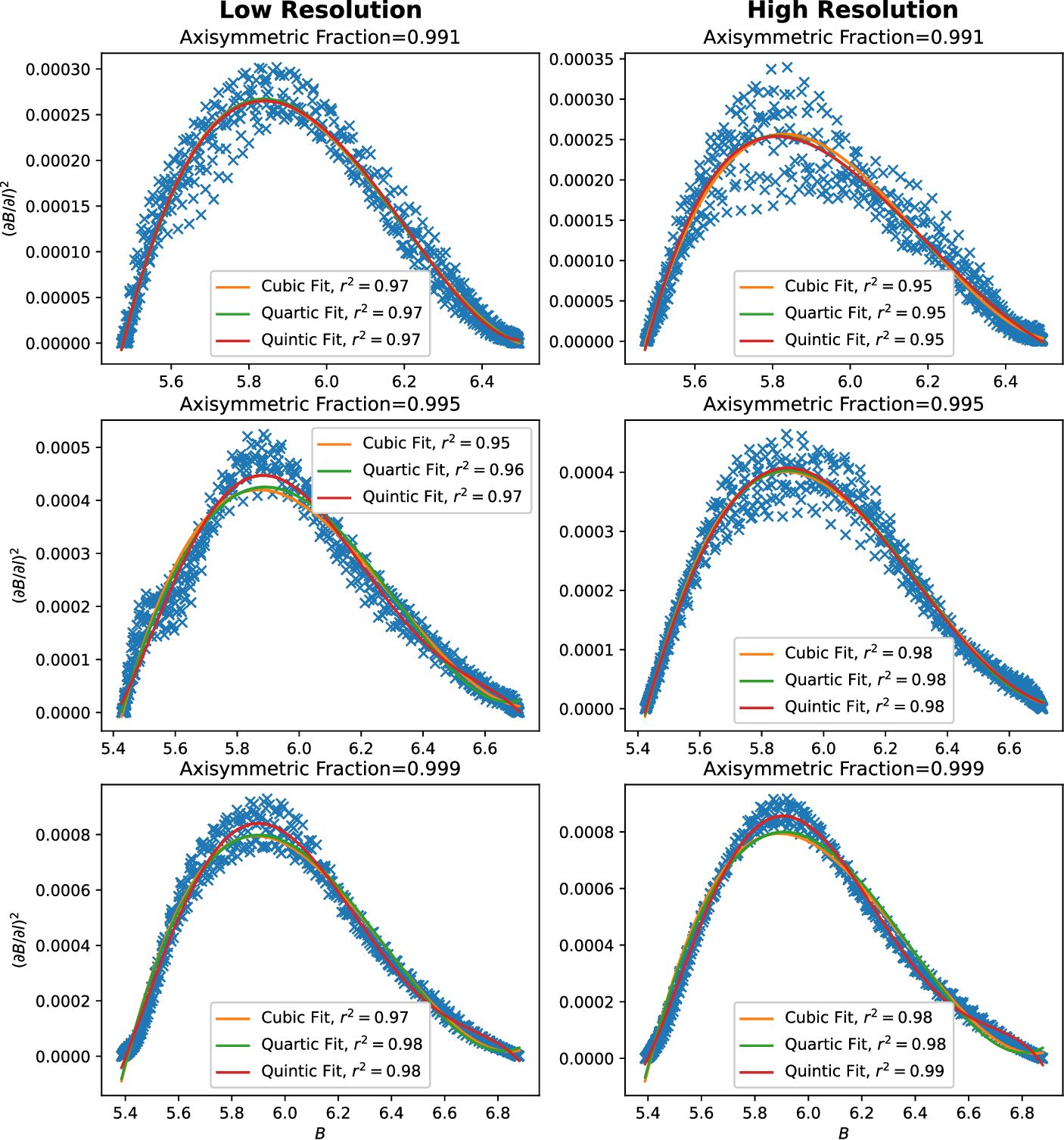}
    \caption{$(\partial_l B)^2$ against $B$ for configurations at three axisymmetric fractions, with their low (first column) and high resolution (second) versions. The low-degree polynomial structure becomes more precise at high resolution.}
    \label{fig:lowres to hires tokamaknessscan}
\end{figure}

We consider two sets of comparisons, as shown in Figure \ref{fig:lowres to hires iotascan} and Figure \ref{fig:lowres to hires tokamaknessscan}. The high-resolution configurations of the former are the same as that considered in the main text and in Section \ref{sec:near_axisym} (cf. Figure \ref{fig:dBdl^2vsB} and Figure \ref{fig:dBdl^2vsB axisymmetric}): they were generated by upscaling their low-resolution cousins, which themselves were generated by optimization. We see that non-polynomial (of low degree) artifacts disappear at higher resolution, with the single exception of the first configuration in Figure \ref{fig:lowres to hires tokamaknessscan}. However, note that this configuration was on the edge of axisymmetric fractions for which {\tt VMEC} would provide a converged solution.

\section*{Author Declarations}

The authors declare that they have no conflicts of interest.

\begin{acknowledgments}
 The authors also thank Per Helander, Michael Maul, Eduardo Rodriguez,  and Rogerio Jorge for fruitful discussions.

This research was supported by a grant from the Simons Foundation/SFARI (560651, AB) and the Department of Energy Award No. DE-SC0024548 (until March 31, 2025). Some computations were performed on the HPC system Viper at the Max Planck Computing and Data Facility (MPCDF).
\end{acknowledgments}

\section*{Data Availability Statement}

The data that support the findings of this study are available from the corresponding author upon reasonable request.

\bibliography{references}% Produces the bibliography via BibTeX.

\end{document}